\documentclass[%
superscriptaddress,
frontmatterverbose,
twocolumn,
reprint
nofootinbib,
 amsmath,
 amssymb,
 aps,
pra,
showpacs
]{revtex4-1}

\usepackage{graphicx}
\usepackage{dcolumn}
\usepackage{bm}
\usepackage[mathlines]{lineno}
\usepackage{physics}
\usepackage{longtable}
\usepackage{siunitx}

\usepackage{color}
\definecolor{linkcol}{rgb}{0,0,0.4}
\definecolor{citecol}{rgb}{0.5,0,0}
\usepackage{hyperref}
\hypersetup
{
pdftitle="Electron Radiated Power in Cyclotron Radiation Emission Spectroscopy Experiments In Waveguides",
bookmarks=true,         
unicode=false,          
pdftoolbar=true,        
pdfmenubar=true,        
pdffitwindow=true,     
pdfstartview={FitH},    
colorlinks=true,       
linkcolor=blue,          
citecolor=red,        
}

\usepackage{environ}
\NewEnviron{comment}{%
  \ifnum\keepcomment=1
    \BODY
  \fi
}
\newcommand{\keepcomment}{0}

\ifnum\keepcomment = 1
	\usepackage{draftwatermark}
    \SetWatermarkColor[rgb]{0.33,0.33,0.33}
	\SetWatermarkLightness{0.85}
	\SetWatermarkText{Project 8 Draft Do Not Distribute}
	\SetWatermarkScale{2}
\fi

\newcommand{\sinc}{\mathrm{sinc}}

\NewEnviron{resize}{
\resizebox{.95\linewidth}{!}{
$\BODY$
 }
 }

\begin{document}


\title{Electron Radiated Power in Cyclotron Radiation Emission Spectroscopy Experiments}

\author{A.~Ashtari Esfahani} \email{ashtari@uw.edu} \affiliation{Center for Experimental Nuclear Physics and Astrophysics and Department of Physics, University of Washington, Seattle, WA 98195, USA}
\author{V.~Bansal} \affiliation{Pacific Northwest National Laboratory, Richland, WA 99354, USA}
\author{S.~B\"oser} \affiliation{Institut f\"ur Physik, Johannes-Gutenberg Universit\"at Mainz, 55128 Mainz, Germany}
\author{N.~Buzinsky} \affiliation{Laboratory for Nuclear Science, Massachusetts Institute of Technology, Cambridge, MA 02139, USA}
\author{R.~Cervantes} \affiliation{Center for Experimental Nuclear Physics and Astrophysics and Department of Physics, University of Washington, Seattle, WA 98195, USA}
\author{C.~Claessens} \affiliation{Institut f\"ur Physik, Johannes-Gutenberg Universit\"at Mainz, 55128 Mainz, Germany}
\author{L.~de~Viveiros} \affiliation{Department of Physics, Pennsylvania State University, State College, PA 16801, USA}
\author{P.~J.~Doe} \affiliation{Center for Experimental Nuclear Physics and Astrophysics and Department of Physics, University of Washington, Seattle, WA 98195, USA}
\author{M.~Fertl} \affiliation{Center for Experimental Nuclear Physics and Astrophysics and Department of Physics, University of Washington, Seattle, WA 98195, USA}
\author{J.~A.~Formaggio} \affiliation{Laboratory for Nuclear Science, Massachusetts Institute of Technology, Cambridge, MA 02139, USA}
\author{L.~Gladstone} \affiliation{Department of Physics, Case Western Reserve University, Cleveland, OH 44106, USA} 
\author{M.~Guigue} \thanks{Now at Sorbonne Universit\'e and Laboratoire de Physique Nucl\'eaire et des Hautes \'Energies, CNRS/IN2P3, 75005 Paris, France} \email{mguigue@lpnhe.in2p3.fr} \affiliation{Pacific Northwest National Laboratory, Richland, WA 99354, USA}
\author{K.~M.~Heeger} \affiliation{Wright Laboratory, Department of Physics, Yale University, New Haven, CT 06520, USA}
\author{J.~Johnston} \affiliation{Laboratory for Nuclear Science, Massachusetts Institute of Technology, Cambridge, MA 02139, USA}
\author{A.~M.~Jones} \affiliation{Pacific Northwest National Laboratory, Richland, WA 99354, USA}
\author{K.~Kazkaz} \affiliation{Lawrence Livermore National Laboratory, Livermore, CA 94550, USA}
\author{B.~H.~LaRoque} \affiliation{Pacific Northwest National Laboratory, Richland, WA 99354, USA}
\author{M.~Leber} \affiliation{Department of Physics, University of California Santa Barbara, CA 93106, USA}
\author{A.~Lindman} \affiliation{Institut f\"ur Physik, Johannes-Gutenberg Universit\"at Mainz, 55128 Mainz, Germany}
\author{E.~Machado} \affiliation{Center for Experimental Nuclear Physics and Astrophysics and Department of Physics, University of Washington, Seattle, WA 98195, USA}
\author{B.~Monreal} \affiliation{Department of Physics, Case Western Reserve University, Cleveland, OH 44106, USA}
\author{E.~C.~Morrison} \affiliation{Pacific Northwest National Laboratory, Richland, WA 99354, USA}
\author{J.~A.~Nikkel} \affiliation{Wright Laboratory, Department of Physics, Yale University, New Haven, CT 06520, USA}
\author{E.~Novitski} \affiliation{Center for Experimental Nuclear Physics and Astrophysics and Department of Physics, University of Washington, Seattle, WA 98195, USA}
\author{N.~S.~Oblath} \affiliation{Pacific Northwest National Laboratory, Richland, WA 99354, USA}
\author{W.~Pettus} \affiliation{Center for Experimental Nuclear Physics and Astrophysics and Department of Physics, University of Washington, Seattle, WA 98195, USA}
\author{R.~G.~H.~Robertson} \affiliation{Center for Experimental Nuclear Physics and Astrophysics and Department of Physics, University of Washington, Seattle, WA 98195, USA}
\author{G.~Rybka} \email{grybka@uw.edu} \affiliation{Center for Experimental Nuclear Physics and Astrophysics and Department of Physics, University of Washington, Seattle, WA 98195, USA}
\author{L.~Salda\~na} \affiliation{Wright Laboratory, Department of Physics, Yale University, New Haven, CT 06520, USA}
\author{V.~Sibille} \affiliation{Laboratory for Nuclear Science, Massachusetts Institute of Technology, Cambridge, MA 02139, USA} 
\author{M.~Schram} \affiliation{Pacific Northwest National Laboratory, Richland, WA 99354, USA}
\author{P.~L.~Slocum} \affiliation{Wright Laboratory, Department of Physics, Yale University, New Haven, CT 06520, USA}
\author{Y-H.~Sun} \affiliation{Department of Physics, Case Western Reserve University, Cleveland, OH 44106, USA} 
\author{J.~R.~Tedeschi} \affiliation{Pacific Northwest National Laboratory, Richland, WA 99354, USA}
\author{T.~Th\"ummler} \affiliation{Institut f\"ur Kernphysik, Karlsruher Institut f\"ur Technologie, 76021 Karlsruhe, Germany}
\author{B.~A.~VanDevender} \affiliation{Pacific Northwest National Laboratory, Richland, WA 99354, USA}
\author{M.~ Wachtendonk} \affiliation{Center for Experimental Nuclear Physics and Astrophysics and Department of Physics, University of Washington, Seattle, WA 98195, USA}
\author{M.~Walter} \affiliation{Institut f\"ur Kernphysik, Karlsruher Institut f\"ur Technologie, 76021 Karlsruhe, Germany}
\author{T.~E.~Weiss} \affiliation{Laboratory for Nuclear Science, Massachusetts Institute of Technology, Cambridge, MA 02139, USA}
\author{T.~Wendler} \affiliation{Department of Physics, Pennsylvania State University, State College, PA 16801, USA} 
\author{E.~Zayas} \affiliation{Laboratory for Nuclear Science, Massachusetts Institute of Technology, Cambridge, MA 02139, USA}

\collaboration{Project 8 Collaboration}
\date{\today}

\begin{abstract}
The recently developed technique of Cyclotron Radiation Emission Spectroscopy (CRES) uses frequency information from the cyclotron motion of an electron in a magnetic bottle to infer its kinetic energy.
Here we derive the expected radio frequency signal from an electron in a waveguide CRES apparatus from first principles.
We demonstrate that the frequency-domain signal is rich in information about the electron's kinematic parameters, and extract a set of measurables that in a suitably designed system are sufficient for disentangling the electron's kinetic energy from the rest of its kinematic features.
This lays the groundwork for high-resolution energy measurements in future CRES experiments, such as the Project 8 neutrino mass measurement.
\end{abstract}

\pacs{29.40.-n, 23.40.-s, 52.50.Qt}
\keywords{Suggested keywords}

\maketitle

\begin{comment}
\listoffigures
\tableofcontents
\newpage
\end{comment}

\begin{comment}
\textbf{Note:} The subsection titles containing a * are here only to help guiding the writer in his story and should be removed once done.
\end{comment}

\section{Introduction to Cyclotron Radiation Emission Spectroscopy }
\label{sec:introduction}
Following the invention of the Penning trap \cite{Penning1936}, low-energy electrons bound with electric and magnetic fields have been used to make some of the most precise measurements of fundamental physics values (e.g., the $g-2$ of the electron \cite{VanDyck1984,Hanneke2008}).  The success of these measurements was contingent on a well-developed theory relating the signal from the axial motion of the electron in the trap to the electron's kinematic parameters \cite{Brown1986}.

Recently it has been proposed \cite{Monreal2009} that the technique called Cyclotron Radiation Emission Spectroscopy (CRES) be used to make precise measurements of the energies of electrons trapped in a magnetic bottle.
The cyclotron radiation from these particles gives direct information about their total energy.
Since an electric field would introduce a position-dependent component to the particle energy, studies of radioactive decay require a purely magnetic trap, eliminating the possibility of a full Penning trap configuration.

Current implementations of CRES \cite{Asner2015,Ashtari2017} consist of electrons produced and trapped inside a waveguide.
The waveguide propagates the cyclotron radiation emitted by the electrons to a receiver with minimal losses.
The background magnetic field within the waveguide consists of two contributions: a strong, uniform, background field, which is parallel to the axis of the waveguide, and a magnetic distortion which forms the magnetic bottle.

Here we develop a mathematical description that relates the characteristics of the apparatus, the motion of the electron, and the measured signal.  
In section \ref{sec:electron-motion-magnetic-bottle}, we investigate the variation of the cyclotron frequency due to the electron's motion.
In section \ref{sec:radiation-trapped-electron}, we derive the electron's radiation spectrum into the waveguide.
In section \ref{sec:effect-experimental-config}, we study the effects of signal reflections on the measured radiation spectrum.
In section \ref{sec:trapping-geometries}, we apply the formulae from sections \ref{sec:electron-motion-magnetic-bottle}, \ref{sec:radiation-trapped-electron}, and \ref{sec:effect-experimental-config} to two examples of magnetic bottle configurations.
Finally, in sections \ref{sec:spectral-features} and \ref{sec:extraction-of-kinematic-parameters}, we demonstrate that there is sufficient information in the signal to reconstruct the kinematic parameters of the electron.

\section{Motion and Cyclotron Frequency of an Electron in A Magnetic Bottle}
\label{sec:electron-motion-magnetic-bottle}

\subsection{The need for a magnetic trap}

The angular cyclotron frequency, $\Omega _c$, of an electron with kinetic energy $K_e$ and mass $m_e$, in a magnetic field $B$, is given by
\begin{equation}
\Omega _c = \frac{eB}{\gamma m_e} = \frac{eB}{m_e + K_e/c^2},
\label{eqn:cyclotron}
\end{equation}
where $e$ is the elementary charge, $c$ is the speed of light, and $\gamma$ the electron's Lorentz factor.
For a known magnetic field, a measurement of the frequency of an electron's cyclotron radiation is also a determination of its kinetic energy \cite{Asner2015}.
The frequency resolution of the measurement, and therefore the energy resolution, improves with increasing observation time. Therefore a no-work trap is necessary for an electron to be observed for a sufficiently long time.

\subsection{Magnetic bottle and pitch angle definition}
A magnetic bottle consists of a local minimum in the magnitude of background magnetic field.
The behavior of a charged particle in a magnetic bottle has been well-described \cite{dehmelt1973proposed,Brown1986}, so here we highlight only the elements important for our results.
If we define an electron's instantaneous pitch angle, $\theta(t)$, as the angle between the local magnetic field and the electron's momentum, then the kinetic energy for an electron undergoing cyclotron motion can be decomposed to its parallel and perpendicular components as
\begin{equation}\label{totenergy}
\begin{split}
K_e & = K_{e\parallel} + K_{e\bot} \\
& = \frac{1}{2}\frac{p_0^2}{m_e}\cos ^2 \theta (t)+\mu (t) B(t),
\end{split}
\end{equation}
where $p_0$ is the magnitude of the electron's initial momentum and $\mu$ is the equivalent magnetic moment of the electron, given by
\begin{equation}\label{eq:eq-magnetic-moment}
\mu (t)=\frac{1}{2}\frac{p_0^2}{m_e}\frac{\sin ^2 \theta (t)}{B(t)}.
\end{equation}

In the adiabatic regime, where the change in the magnetic field direction is slow compared with the cyclotron frequency, an electron's equivalent magnetic moment is a constant of motion.
For the remainder of this derivation $\mu$ is treated as time-independent and the term $\mu B(t)$ behaves as a magnetic potential energy.
Electrons with pitch angles of $90 + \delta\theta$ or $90 - \delta\theta$ degrees will have the same motion; therefore we will only consider electrons with pitch angles between 0 and \SI{90}{degrees}.
The pitch angle approaches \SI{90}{degrees} for an electron exploring regions of increasing magnetic field, whereas the pitch angle decreases for an electron approaching the bottom of the trap.
For every electron, we define the pitch angle at the bottom of the trap to be $\theta_{bot}$.
Due to conservation of energy, the condition on pitch angle for a trapped electron is
\begin{equation}\label{eq:trapped-pitch-angles}
\theta_{bot} \geq \sin^{-1} \left(\sqrt{1-\frac{\Delta B}{B_{\mathrm{max}}}}\right),
\end{equation}
where $B_{\mathrm{max}}$ is the maximum value of the magnetic field and $\Delta B$ is the trap depth.
\begin{figure}
\includegraphics[width=0.48\textwidth]{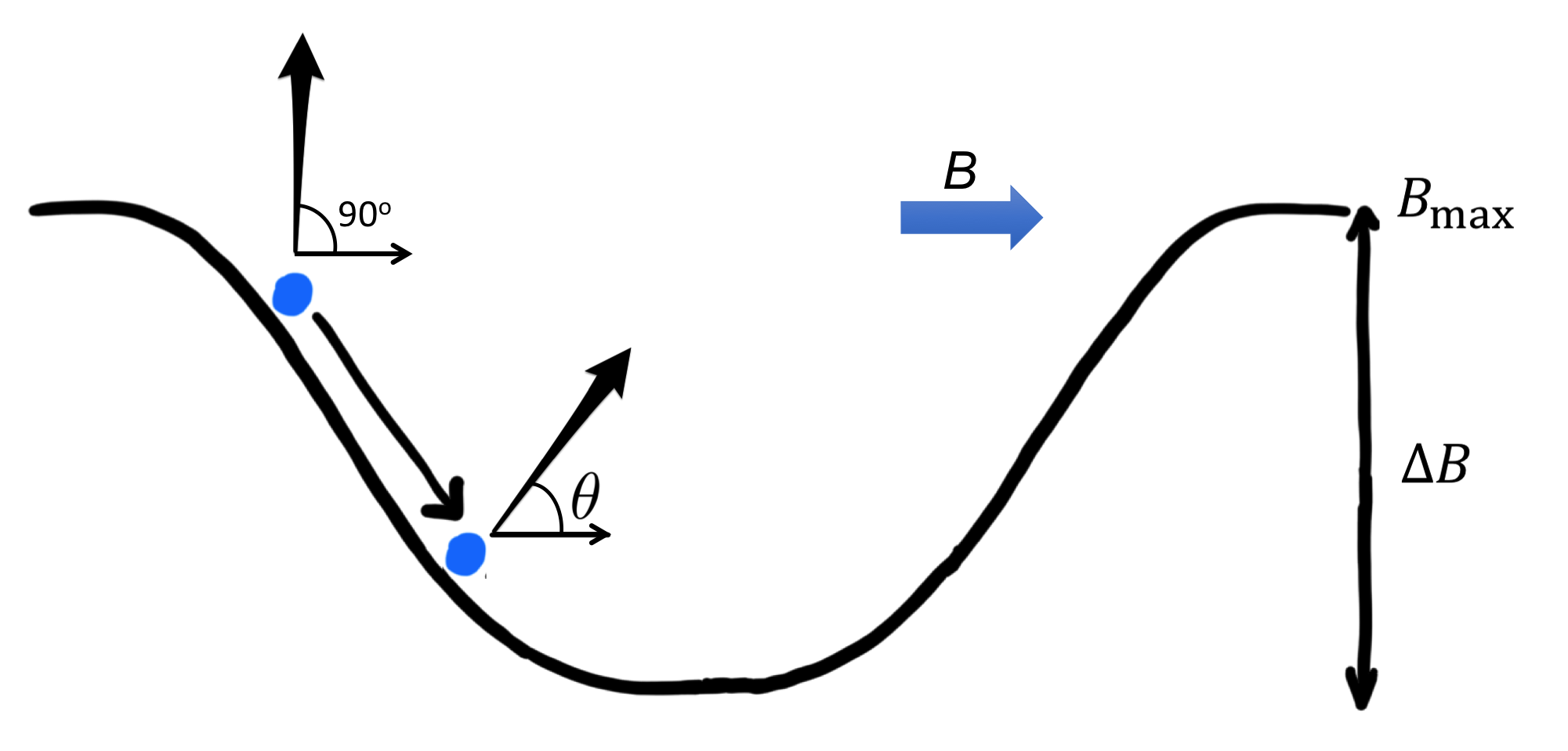}
\caption{\label{fig:motion} Axial motion of an electron in a magnetic bottle.
The magnetic trap has depth $\Delta B$ and maximum value $B_{max}$.
The electron's pitch angle is defined as the angle between the electron's momentum vector and the direction of the local magnetic field.
If the electron's pitch angle at the bottom of the trap satisfies Eq. \eqref{eq:trapped-pitch-angles}, the electron undergoes an oscillatory axial motion inside the trap.
The turning point for the electron corresponds to the position when the pitch angle is 90 degrees.
}
\end{figure}

Existing CRES experiments operate at a background field of \SI{1}{T}, which we use for all examples throughout.
A \SI{4}{mT} trap depth on this background field can trap electrons with pitch angles greater than \SI{86}{degrees}.
As a consequence, electrons trapped in the magnetic bottle with a pitch angle other than \SI{90}{degrees} at the bottom of the trap will undergo periodic axial motion as depicted in Fig. \ref{fig:motion}.

\subsection{Time-varying cyclotron frequency of an electron}

The magnetic field experienced by an electron varies with time due to its axial motion, resulting in a time-varying cyclotron frequency, given by
\begin{equation}\label{eqn:time-varying-cyclotron}
\Omega _c(t) = \frac{eB(t)}{m_e + K_e/c^2}.
\end{equation}

Additionally, the electron's cyclotron motion causes it to radiate, reducing its kinetic energy and therefore increasing its cyclotron frequency.
This energy loss can be expressed as
\begin{equation}
\frac{\mathrm{d}K_e (t)}{\mathrm{d}t} =-P(t),
\end{equation}
where  $P$, the power radiated by the electron, can be assumed to be constant over short times.
The energy radiated is much smaller than the electron’s initial total energy.
Therefore, the instantaneous frequency of radiation emitted by the electron can be derived from Eq. \eqref{eqn:cyclotron} as
\begin{equation}
\Omega_c(t) \simeq \frac{eB(t)}{m_e+K_0/c^2} \left(1+ \frac{Pt}{m_ec^2 + K_0}\right),
\label{eqn:approx_instantaneous_cyclotron}
\end{equation}
where $K_0$ is the initial kinetic energy of the electron.
Because the cyclotron radiation is observed for a finite amount of time, the frequency is shifting by the electron power loss $\frac{Pt}{m_ec^2 + K_0}$.

Existing CRES experiments operate with midly relativistic electrons, we take a $30~\mathrm{keV}$ electron for examples throughout.
Such an electron in a \SI{1}{T} background field radiates $1~\mathrm{fW}$ of power.
Over $10~\mathrm{\mu s}$ this results in a cyclotron frequency shift of $3~\mathrm{kHz}$, which is equivalent to an energy shift of $60~\mathrm{meV}$.
This effect can be ignored in the following calculation of CRES power spectral density.
We will consider it again when we introduce the slope of tracks in Sec. \ref{sec:spectral-features}.

\subsection{Axial motion and Doppler shift\label{sec:axial-motion}}

As a trapped electron oscillates axially in a magnetic bottle, the frequency of radiation collected by the receiver on the same axis, $\Omega _r$,  is shifted by the Doppler effect and can be expressed as
\begin{equation}\label{eq:doppler-shifted-frequency}
\Omega_r(t) = \Omega_c(t_{ret}) \times \left(1-\frac{v_z(t_{ret})}{v_p}\right)^{-1},
\end{equation}
where $t_{ret}$ is the retarded time, $v_z$ is the electron axial velocity, and $v_p$ is the phase velocity of the wave inside the waveguide.
For mildly relativistic electrons with the large pitch angles required for trapping, the term $\frac{v_z(t_{ret})}{v_p}$ is small compared to 1.
Substituting $\Omega_c$ from Eq. \eqref{eqn:approx_instantaneous_cyclotron} into Eq. \eqref{eqn:time-varying-cyclotron} results in
\begin{equation}\label{eq:simple-doppler-shifted-frequency}
\begin{split}
\Omega_r(t) &\simeq \frac{eB(t_{ret})}{m_e+K_0/c^2}  \left( 1 +\frac{v_z(t_{ret})}{v_p} \right),
\end{split}
\end{equation}
in which the second-order  contributions in $v_z/v_p$ have been neglected.
Eq. \eqref{eq:simple-doppler-shifted-frequency} introduces two systematic effects that must be accounted for to understand the relationship between the electron's energy and the observed signal.

First, the average value of $B(t)$ is greater than the value of $B$ at the center of the trap and depends on the magnitude of the electron's axial motion.
This causes the average measured frequency to be dependent on the electron's motion in the trap.
This feature has been briefly discussed in \cite{Asner2015}, for electrons in a harmonic trap, and will be discussed in detail in Section \ref{sec:trapping-geometries}.

\begin{figure}
\includegraphics[width=0.48\textwidth]{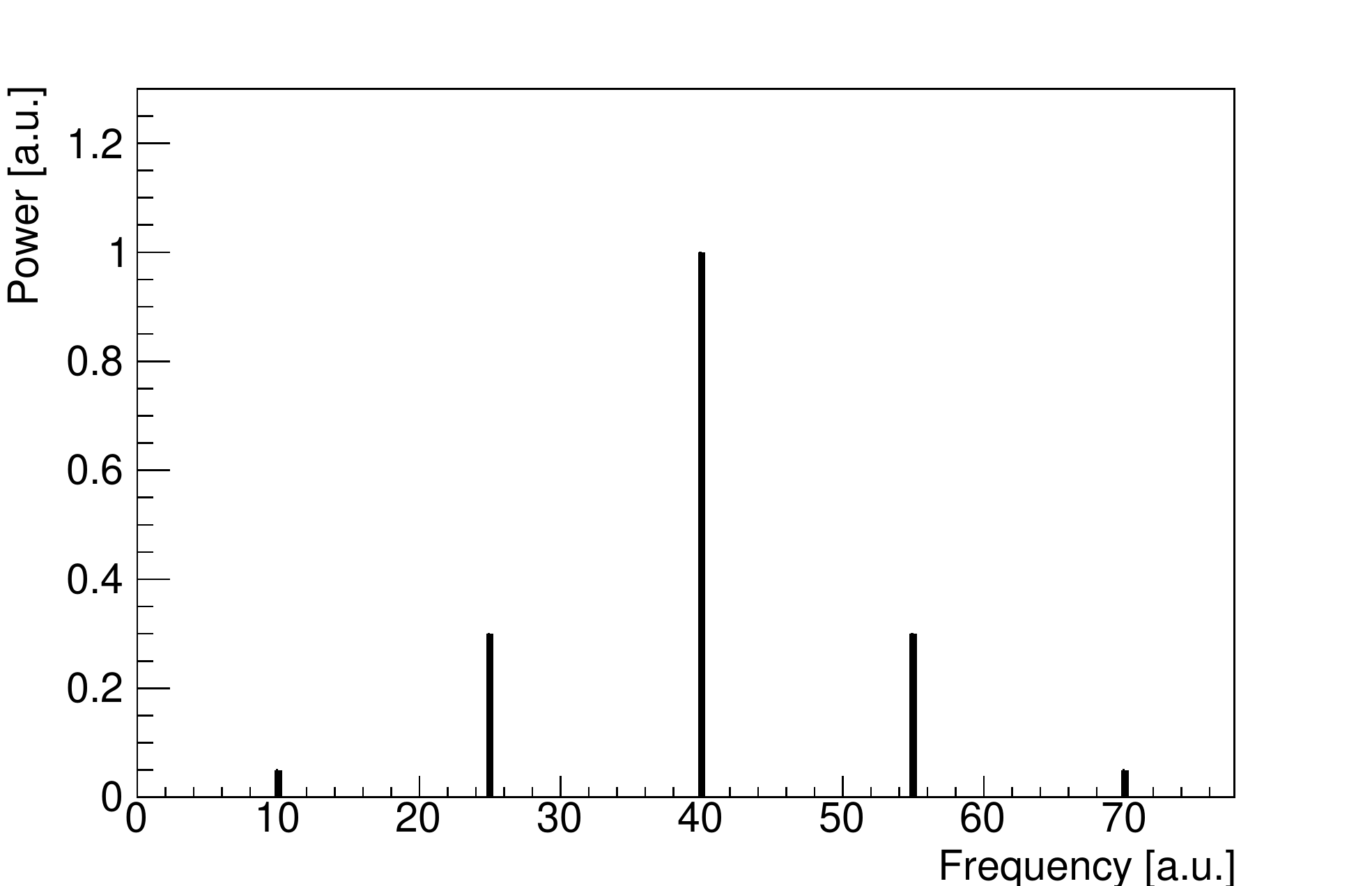}
\caption{\label{fig:comb-structure}
The comb structure of the frequency spectrum of cyclotron power from a trapped \SI{30}{keV} electron in a \SI{1}{T} background field.
The central peak is located at the average cyclotron frequency, and the axial frequency, which defines the separation between the peaks, is $15~\mathrm{MHz}$.
}
\end{figure}

Second, the terms $B(t)$ and $v_z(t)$ vary periodically at harmonics of the frequency of the electron's axial motion.
This imposes frequency modulation on the cyclotron signal, both by the varying magnetic field and by the Doppler shift, with the modulation due to the magnetic field being the smaller of the two effects.
Frequency modulated signals have been studied extensively as a form of encoding information in radio frequency signals \cite{Sekhar2005}.
The expected signal at the receiver consists of a frequency comb structure, where the main carrier is at the average cyclotron frequency and is surrounded by sidebands which are evenly spaced by the frequency of axial motion as shown in Fig.~\ref{fig:comb-structure}.

The relative magnitude of the sidebands can be characterized by the modulation index, $h=\frac{\Delta \omega}{\omega_a}$, where $\Delta \omega$ is the maximum frequency change due to the Doppler shift and magnetic field, and $\omega_a$ is the axial frequency.  The magnitude of the $n^{\mathrm{th}}$ sideband is given by the Bessel function $J_n\left(h\right)$.
For values of $h$ greater than 0.5, a significant fraction of power is present in the sidebands.
For $h\simeq2.41$,  all of the power is radiated in sidebands, and no power is radiated into the carrier, as shown in Fig. \ref{fig:sideband_magnitude}.

As a simple case, we can calculate the sideband structure from only the Doppler shift for an electron moving axially in simple harmonic motion, with axial frequency $\omega_a$ and maximum travel $z_{\mathrm{max}}$.
From Eq. \eqref{eq:simple-doppler-shifted-frequency} we note that the maximum frequency change is $\Delta \omega=\Omega_c \omega_a z_{\mathrm{max}} / v_p$.
The modulation index is then $h=\frac{\Omega_c z_{\mathrm{max}}}{v_p}$.
The threshold for significant received signal power in the sidebands, $h \sim 0.5$, is therefore equivalent to an axial travel for electrons greater than a half-wavelength of light in waveguide.

\subsection{Grad-B and Curvature Drifts \label{sec:grad-b-motion}}

The electron undergoes a cyclotron motion, an axial motion, and two drift motions induced by non-uniformity in the magnetic field.
The first force is prompted by the magnetic gradient in the trap.
These local magnetic field gradients exert a force on the electron that gives rise to a drift velocity perpendicular to both the magnetic field and its gradient, which we call grad-B motion, given by \cite{Northrop}
\begin{equation}\label{eq:grad-b}
\mathbf{v}_{grad-B} = \frac{\mu}{m\Omega_cB} \mathbf{B} \times \nabla B.
\end{equation}

This slow grad-B motion is analogous in its effect to magnetron motion in a Penning trap; it pins the guiding center of the electron’s cyclotron motion to a larger circle. The radius of this larger circle is set by the electron’s radial position in the trap at the moment it is created.

The grad-B velocity for a \SI{30}{keV} electron with pitch angle of \SI{86}{degrees} in a \SI{1}{T} magnetic field with a $10~\mathrm{mT/m}$ field gradient is smaller than $300~\mathrm{m/s}$.
This velocity corresponds to a frequency of \SI{5}{kHz} for an electron orbit with a \SI{1}{cm} radius.
For power spectral densities calculated for a finite time length smaller than grad-B motion's period, this effect can be ignored and the electron's guiding center can be assumed fixed.

The curvature in the field lines introduces another drift motion, which we call curvature drift, given by \cite{Northrop}
\begin{equation}\label{eq:curv}
\mathbf{v}_{curv} = \frac{v_0^2 \cos^2(\theta(t))}{\Omega_cB^3} \mathbf{B} \times \left(\mathbf{B} \cdot \nabla\right) \mathbf{B}.
\end{equation}
For the conditions described  below Eq.\eqref{eq:grad-b}, the curvature drift is smaller than $3~\mathrm{m/s}$ and therefore negligible. For a detailed study of these two effects look at \cite{Fitzpatrick2014} and \cite{Jackson}.


\section{Radiation of a Trapped Electron into a Waveguide Mode}
\label{sec:radiation-trapped-electron}

We now derive generic expressions for the spectral distribution of the cyclotron radiation of a trapped electron.
We first expand the radiation from a generic current inside the waveguide volume in terms of waveguide modes, and derive the power that propagates through the waveguide.
We then discuss the specific case of an electron coupling to a waveguide and the associated approximations (as done with more detail in \cite{Collin1965}).
This allows us to show that the mode excitation can be written in term of harmonics, corresponding to the axial modes, which demonstrates the comb structure of the measured cyclotron power.
Finally, we discuss the implications of our results in two examples, rectangular and circular waveguide.

\subsection{Waveguide modes and transmitted power}

Generalizing the notation in Jackson, \cite{Jackson}, the  electric and magnetic fields inside a waveguide can be written as a sum over all modes  in the $\pm$ z-directions as
\begin{equation}\label{eq3.1}
\begin{resize}
\begin{split}
\textbf{E}^{\pm}(\mathbf{r},t) &= \sum_\lambda \int _{-\infty}^\infty A_{\lambda}^\pm(\omega) \left(\textbf{E}_{t\lambda}(x,y)  \pm E_{z\lambda}(x,y) \textbf{\^{z}}\right)  \\
&\times e^{\pm i k_\lambda z} e^{-i\omega t} \mathrm{d}\omega\\
\textbf{H}^{\pm}(\mathbf{r},t) &= \sum_\lambda \int _{-\infty}^\infty A_{\lambda}^\pm(\omega) \left(\pm \frac{1}{Z_\lambda}\textbf{\^{z}}\times\textbf{E}_{t\lambda}(x,y) + H_{z\lambda}(x,y) \textbf{\^{z}}\right)  \\
&\times e^{\pm i k_\lambda z} e^{-i\omega t} \mathrm{d}\omega,
\end{split}
\end{resize}
\end{equation}
where $k_\lambda$ is the wave number and $Z_{\lambda}$ the mode impedance.
The amplitude, $A_{\lambda}(\omega)$, of each mode is found via Poynting's theorem and is given by
\begin{equation}\label{eq:amplitude-definition}
A_{\lambda}^\pm(\omega)  = -\frac{Z_\lambda}{2} \int _V \textbf{J}(\omega)\cdot\left(\textbf{E}_{t\lambda}(x,y)  \mp E_{z\lambda}(x,y) \textbf{\^{z}}\right) e^{\mp i k_{\lambda}z}\mathrm{d}^3 r,
\end{equation}
with $V$ being the waveguide volume and the current inside the waveguide, $\textbf{J}(\omega)$, defined as
\begin{equation}\label{eq:current-density}
\textbf{J}(\omega) = \frac{1}{2\pi} \int _{-\infty}^\infty \textbf{J}(\textbf{r},t) e^{i\omega t} \mathrm{d}t .
\end{equation}
These mode amplitudes fully determine the signal in the waveguide.
The transverse electric field modes, $\textbf{E}_{t\lambda}(x,y)$, are normalized over the waveguide cross-section $\mathcal{A}$ such that
\begin{eqnarray}\label{eq3.2}
\int _{\mathcal{A}} \textbf{E}_{t\lambda} \cdot \textbf{E}_{t\mu} \mathrm{d} a = \delta_{\lambda \mu}.
\end{eqnarray}
The longitudinal electric field modes, $\textbf{E}_{z\lambda}(x,y)$, are normalized for TM modes such that
\begin{equation}\label{eq:long-field-normalization}
\int _{\mathcal{A}} E_{z\lambda} \cdot E_{z\mu} \mathrm{d} a = -\frac{\gamma_\lambda^2}{k_\lambda^2}\delta_{\lambda \mu},
\end{equation}
with $\gamma_\lambda$ being the mode eigenvalues, which are zero for TE modes.

The power transmitted in the $\pm$ z-direction is a spatial integral of the normal component of the Poynting vector, taken over the waveguide's cross section, $\mathcal{A}$.
It can be written as
\begin{equation}\label{eq:poynting-along-z}
P^\pm(t) = \int_{\mathcal{A}} \textbf{E}^\pm(\mathbf{r},t) \times \textbf{H}^{\pm} (\mathbf{r},t) \cdot (\pm \textbf{\^{z}}) \,\mathrm{d}a = \sum_\lambda \frac{1}{Z_\lambda} \left[ B_\lambda^\pm(t)\right]^2,
\end{equation}
in which the mode excitation, $B_\lambda^\pm(t)$, not to be confused with the B-field, is defined as
\begin{equation}\label{eq:B_amplitude_definition}
B_\lambda^\pm(t) = \int _{-\infty}^\infty A_\lambda^\pm(\omega) e^{\pm i k_{\lambda} z} e^{-i\omega t} \mathrm{d} \omega .
\end{equation}

\subsection{Power spectral density}

Power spectral density is the quantity which we ultimately aim to calculate.
To that end, we define the power spectral density of the mode excitation as
\begin{equation}\label{eq:power-spectral-density}
{\tilde{P}}^\pm(\omega) = \frac{2\pi}{T} \sum_\lambda \frac{1}{Z_\lambda} \left\vert{\tilde{B}}_\lambda^\pm(\omega)\right\vert^2,
\end{equation}
with
\begin{equation}\label{eq:spectral-amplitude-zprop}
\begin{split}
{\tilde{B}}^{\pm}_\lambda(\omega) & = \frac{1}{2\pi}\int_{-\infty}^{+\infty} B^{\pm}_\lambda (t) e^{i\omega t}\mathrm{d}t= A_\lambda^\pm(\omega) e^{\pm i k_{\lambda} z},
\end{split}
\end{equation}
and $T$ being the total time of observation.
Eq. \eqref{eq:power-spectral-density} can be interpreted as the sum of the power in each waveguide mode,
\begin{equation}\label{eq:sum-power}
{\tilde{P}}^\pm(\omega) = \sum _{\lambda} \tilde{P}_{\lambda}^\pm (\omega),
\end{equation}
where
\begin{equation}\label{eq:mode_power}
\tilde{P}_{\lambda}^\pm (\omega)=  \frac{2\pi}{T} \frac{1}{Z_\lambda} \left\vert {\tilde{B}}_\lambda^\pm(\omega) \right\vert^2 .
\end{equation}

In the case of a single electron, moving on the trajectory $\textbf{r}=\textbf{r}_0(t)$ with the velocity $\textbf{v}(t)$, the current density is
\begin{equation}\label{eq:current-single-electron}
\textbf{J}(\textbf{r},t) = - e \textbf{v}(t) \delta^3(\textbf{r}-\textbf{r}_0(t)).
\end{equation}
From this and Eq. \eqref{eq:amplitude-definition}, the mode amplitudes can be found to be
\begin{equation}\label{eq:amplitude-decomposed}
\begin{split}
A_{\lambda}^\pm(\omega)  = -\frac{Z_\lambda}{4\pi} \int _V \int_{-\infty}^{\infty} & e \textbf{v}(t) \cdot\left(\textbf{E}_{t\lambda}(x,y)  \mp E_{z\lambda}(x,y) \textbf{\^{z}}\right) \\
&\times \delta^3(\textbf{r}-\textbf{r}_0(t)) e^{i\omega t}e^{\mp ik_{\lambda}z}\mathrm{d}t\mathrm{d}^3 r.
\end{split}
\end{equation}
By changing the order of integrals and taking the spatial integral, we find that
\begin{equation}\label{eq:field_amplitudes}
\begin{split}
A_{\lambda}^\pm(\omega)  & = -\frac{e Z_\lambda}{4\pi} \int_{-\infty}^{\infty} \textbf{v}(t) \cdot\left[\textbf{E}_{t\lambda}(x_0(t),y_0(t))  \right.\\
&\left.\mp E_{z\lambda}(x_0(t),y_0(t)) \textbf{\^{z}}\right] e^{i\omega t}e^{\mp ik_{\lambda}z_0(t)}\mathrm{d}t,
\end{split}
\end{equation}
where the field is evaluated at the electron's position, $\textbf{r$_0(t)$}=(x_0(t),y_0(t),z_0(t))$.
Using the mode amplitudes, the procedure from the preceding section is used to find ${\tilde{B}}^{\pm}_\lambda(\omega)$, from which the energy losses and signal power follow.

\subsection{Field amplitudes for a CRES electron}

Eq. \eqref{eq:field_amplitudes} describes the coupling of an electron inside a waveguide, without any assumptions about its motion.
A number of reasonable approximations can be used in the case of an electron in a CRES experiment.

The electron's periodic motion can be decomposed into a cyclotron motion, an axial motion and a drift motion.
Following the discussion of section \ref{sec:grad-b-motion} we assume this last motion is slow compared with the first two, so the electron's transverse and longitudinal velocity components can be written as
\begin{equation}\label{eq:transverse-velocity}
\begin{split}
\textbf{v}_t (t) &= v_0 \sin \theta (t) \left(  \cos \Phi _c (t) \textbf{e}_1   + \sin \Phi _c (t) \textbf{e}_2 \right) \\
v_z (t) &= v_0 \cos \theta (t),
\end{split}
\end{equation}
where $(\textbf{e}_1,\textbf{e}_2)$ is an orthonormal basis in the plane transverse to the $z$ direction, $v_0$ is the electron's initial velocity, and $\Phi_c(t)$ is the phase of the electron in its cyclotron orbit, defined as
\begin{equation}\label{eq:cyclotron-phase}
\Phi _c (t) = \int _0 ^t \Omega _c (t')\mathrm{d}t'.
\end{equation}
This phase can also be written as a combination of constant phase progression at the average cyclotron frequency, $\Omega _0$, and a periodic perturbation at the electron's axial frequency.

The $\textbf{v}\cdot \textbf{E}$ term in Eq.\eqref{eq:field_amplitudes}, at the position $(x_0(t),y_0(t))$, can then be written as
\begin{equation}\label{eq:amplitude-pitchangle}
\begin{split}
&{\textbf{v}(t)} \cdot\left(\textbf{E}_{t\lambda}  \mp E_{z\lambda} \textbf{\^{z}}\right) = \\
& v_0\sin\theta(t)\left( E_{1\lambda} \cos(\Phi_c(t))+ E_{2\lambda} \sin(\Phi_c(t))\right)  \mp \cos \theta (t) E_{z\lambda}),
\end{split}
\end{equation}
where $E_{1\lambda}$ and $E_{2\lambda}$ are the components of the transverse electric field for the mode $\lambda$.

The radius of the cyclotron motion, $r_c$, and the wavelength of the cyclotron radiation, $\lambda_c$, are related via
\begin{equation}
r_c = \frac{v}{2\pi c} \lambda_c.
\end{equation}
As a result, the radius of cyclotron motion is small compared to the wavelength of cyclotron radiation and therefore the waveguide dimensions.
The variation in coupling due to the cyclotron motion can be neglected, and one can replace the actual position of the electron by its gyrocenter, defined to be the center of the electron's cyclotron motion.
In this work, we will further assume the transverse position of the electron's  gyrocenter $(x_c,y_c)$ does not change with time.
This may not be true in experiments with significant drift motion.

The $z$ component of the $\textbf{v}\cdot \textbf{E}$ term in Eq. \eqref{eq:field_amplitudes} is equal to zero for Transverse Electric (TE) modes and small in Transverse Magnetic (TM) modes for electrons with large pitch angles.
The phase oscillation induced by $\sin\theta (t)$ in Eq. \eqref{eq:amplitude-pitchangle} is thus small compared with the cyclotron phase $\Phi _c$ and can be neglected.

Using the above approximations, Eq. \eqref{eq:amplitude-pitchangle} can be rewritten as
\begin{equation}\label{eq:amplitude-pitchangle-simplified}
\begin{split}
\textbf{v}\cdot \textbf{E} = v_0 &\left[ E_{1\lambda}(x,y) \cos(\Phi_c(t)) + E_{2\lambda}(x,y) \sin(\Phi_c(t))\right]\\
=\frac{v_0}{2}&\left[ (E_{1\lambda}-iE_{2\lambda})e^{i\Phi_c(t)} + (E_{1\lambda}+iE_{2\lambda})e^{-i\Phi_c(t)} \right] .
\end{split}
\end{equation}
Replacing the above expression for $\textbf{v}\cdot \textbf{E}$ in Eq. \eqref{eq:field_amplitudes} we get
\begin{equation}\label{eq:amplitude-series}
\begin{split}
A_{\lambda}^\pm(\omega) & = -\frac{eZ_\lambda v_0}{8\pi} \left[(E_{1\lambda}-i E_{2\lambda})\int _{-\infty}^\infty e^{i\Phi_c(t)} e^{\mp ik_{\lambda}z_0(t)}e^{i\omega t}\mathrm{d}t \right.\\
&\left.+ (E_{1\lambda}+i E_{2\lambda})\int _{-\infty}^\infty e^{-i\Phi_c(t)} e^{\mp ik_{\lambda}z_0(t)}e^{i\omega t}\mathrm{d}t\right] ,
\end{split}
\end{equation}
where the electric fields are being evaluated at the electron's gyrocenter $(x_c,y_c)$.

\subsection{Mode expansion of motion and phase}

Because $z_0(t)$ and $\Phi_c(t)-\Omega_0 t$ are periodic at the electron's axial motion frequency $\Omega _a$, these terms can be expanded in a Fourier series  as
\begin{equation}\label{eq:bathtub-first-expansion}
e^{i\Phi_c(t)-i\Omega_0 t} = \sum_{m=-\infty} ^\infty \alpha_m e^{im\Omega_a t}
\end{equation}
and
\begin{equation}\label{eq:bathtub-second-expansion}
e^{ik_\lambda z(t)} = \sum_{m=-\infty} ^\infty \beta_m(k_\lambda) e^{im\Omega_a t}.
\end{equation}
As a result, the exponential term in Eq. \eqref{eq:amplitude-series} can be written as
\begin{equation}\label{eq:fourier-series}
e^{i\Phi_c(t)+ ik_{\lambda}z_0(t)} = \sum _{n=-\infty} ^{\infty} a_n(k_{\lambda}) e^{i(\Omega_0+n\Omega_a)t},
\end{equation}
in which
\begin{equation}\label{fourier-coefficient}
a_n(k_\lambda) = \sum_{m=-\infty} ^\infty \alpha_m (k_\lambda) \beta_{n-m}(k_\lambda) .
\end{equation}
These coefficients, $a_n$, can be computed from a decomposition of the axial motion and the cyclotron phase evolution into harmonics of the axial frequency.
This greatly simplifies the study of the radiated power spectral density.

Based on Eq. \eqref{fourier-coefficient}, we get the following:
\begin{equation}\label{eq:other-fourier-series}
\begin{split}
e^{i\Phi_c(t)- ik_{\lambda}z_0(t)} & = \sum _{n=-\infty} ^{\infty} a_n(-k_{\lambda}) e^{i(\Omega_0+n\Omega_a)t}, \\
e^{-i\Phi_c(t)- ik_{\lambda}z_0(t)} & = \sum _{n=-\infty} ^{\infty} a_n^*(k_{\lambda}) e^{-i(\Omega_0+n\Omega_a)t}, \\
e^{-i\Phi_c(t)+ ik_{\lambda}z_0(t)} & = \sum _{n=-\infty} ^{\infty} a_n^*(-k_{\lambda}) e^{-i(\Omega_0+n\Omega_a)t} .
\end{split}
\end{equation}
Expanding the exponential terms in Eq. \eqref{eq:amplitude-series} using the above Fourier series results in
\begin{equation}\label{eq:field-amplitude-2}
\begin{resize}
\begin{split}
A_{\lambda}^\pm(\omega) & = -\frac{eZ_\lambda v_0}{2} \left[(E_{1\lambda}-i E_{2\lambda})\sum _{n=-\infty} ^{\infty} a_n(\mp k_{\lambda}) \delta(\omega+\Omega_0+n\Omega_a) \right.\\
&\left.+ (E_{1\lambda}+i E_{2\lambda})\sum _{n=-\infty} ^{\infty} a_n^*(\pm k_{\lambda}) \delta(\omega-\Omega_0-n\Omega_a)\right] .
\end{split}
\end{resize}
\end{equation}

\subsection{Frequency comb structure of cyclotron power}

Utilizing conventional techniques of handling $\delta^2$ functions and the relationship between the wave-number and frequency, the power spectral density for the waveguide mode $\lambda$, Eq. \eqref{eq:mode_power} is
\begin{equation}\label{eq:power_comb_structure}
\begin{split}
{\tilde{P}}^\pm_\lambda(\omega) & = P_{0,\lambda} \sum _{n=-\infty} ^{\infty} \left|a_n \left(\pm \frac{\Omega_0+n\Omega_a}{v_{p,\lambda}}\right)\right|^2  \\
&\times \left[\delta(\omega - (\Omega_0+n\Omega_a)) + \delta(\omega + \Omega_0+n\Omega_a)\right] ,
\end{split}
\end{equation}
where $P_{0,\lambda}$ is defined as
\begin{equation}\label{p0-def}
P_{0,\lambda} = \frac{e^2 v_0^2Z_{\lambda}}{8} \left[E_{1\lambda}^2+ E_{2\lambda}^2\right],
\end{equation}
and $v_{p,\lambda}$ is the phase velocity in the waveguide for the mode $\lambda$.
Note that there are possible cross terms between the $n^{\mathrm{th}}$  positive and the $m^{\mathrm{th}}$ negative frequencies when $n+m = -\frac{2\Omega_0}{\Omega_a}$.
Because of the small values of $a_n$ for large $n$, these terms can be neglected.

The measured power spectrum thus exhibits a comb structure in the frequency domain as shown in Fig.~\ref{fig:comb-structure}.
For an electron with no axial motion, all the power will be radiated with a frequency $\Omega_0$.
An electron with pitch angle other than \SI{90}{degrees} at the bottom of the trap, will undergo axial motion, and as a result some power will be radiated at the harmonic frequencies which are $n \Omega_a$ away from the main peak.
Eq. \eqref{eq:power_comb_structure} indicates that the power in the $n^{\mathrm{th}}$ harmonic is
\begin{equation}\label{eq:sideband-power}
P_n = P_{0,\lambda} \left\vert a_n \left( \pm \frac{\Omega_0+n\Omega_a}{v_{p,\lambda}} \right)\right\vert^2.
\end{equation}

\subsection{Power in particular waveguide geometries}
\label{sec:power-waveguide-geometries}

The simplest experimental design choice is a waveguide geometry in which the radiation from the electron will only couple significantly to a single propagating mode.
Detailed calculations of $P_{0,\lambda}$ for two interesting examples are included in Appx. \ref{sec:power-term}.
For the $TE_{10}$ mode in a rectangular waveguide we get
\begin{equation}\label{eq3.14}
P_{0,TE_{10}} = \frac{Z_{10}e^2v_0^2}{4 wh} \cos^2 \left(\frac{\pi x_c}{w}\right) ,
\end{equation}
in which $Z_{10}$ is the $TE_{10}$ mode impedance, $v_0$ is the electron velocity, $w$ and $h$ are the waveguide's width and height, defined to be along x and y directions respectively, and $x_c$ is the x position of the electron's gyrocenter.

For the $TE_{11}$ mode in a circular waveguide we get
\begin{equation}\label{eq3.15}
\begin{split}
P_{0,TE_{11}} & = \frac{Z_{{11}}e^2 v_0^2}{8\pi \alpha} \left( J_1'^2(k_c \rho_c) + \frac{1}{k_c^2 \rho_c^2} J_1^2(k_c \rho_c)\right) ,
\end{split}
\end{equation}
in which $Z_{11}$ is the $TE_{11}$ mode impedance, $\rho_c$ defines the radial position of the gyrocenter of the electron in cylindrical coordinates, $k_c$ is the wavenumber for the cutoff frequency of the mode, and $\alpha$ is defined in Eq. \eqref{eqa.8}.


\section{Effects of Waveguide Reflection}
\label{sec:effect-experimental-config}

In our discussion of waveguides we have assumed infinite length, whereas any experimental realization of a CRES experiment must be finite in length.  Allowing that one end of the waveguide must have a receiver, we are left with several options for the treatment of signals at the other end.

\begin{figure}
\includegraphics[width=0.48\textwidth]{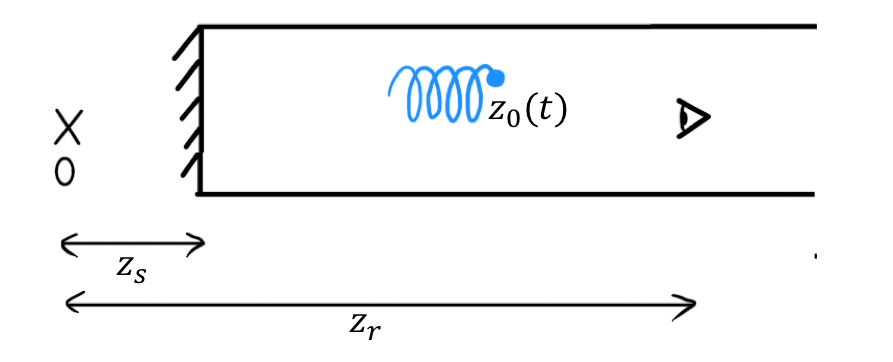}
\caption{\label{fig:CRES} Schematic of an experiment with an electron undergoing cyclotron motion in a waveguide with a conductive short. The relevant parameters include: the origin, $O$, the position of the electron, $z_0$, the position of the waveguide short, $z_s$, on the left side of the waveguide, and the position of the receiver, $z_r$, on the right side of the waveguide.
The magnetic field, $B$, is parallel to the waveguide axis.}
\end{figure}

One option is to add a second receiver.
The signal observed by each receiver is then available for analysis, at the cost of supporting two receiver systems.
Another option is to install a terminator on one end of the waveguide.
The receiver will detect only half of the electron's radiated power and the signal will be the same as the case of the infinite waveguide.
The final option, shown in Fig. \ref{fig:CRES}, is to install a conductive short to the end of the waveguide, reflecting signals back to the first receiver.
The first two options have been already analyzed.
In this section we calculate the effects of the reflector on the power spectral density of the CRES signal.
\begin{comment}
\subsection{*Calculating field amplitude for the interference}
\end{comment}

The total mode excitation at the receiver, $\tilde{B}_{\lambda}(\omega )$, is a superposition of the direct wave, $\tilde{B}^+_{\lambda} (\omega )$, and the reflected wave, $\tilde{B}^-_{\lambda} (\omega )$.
The reflection induces a phase shift of \SI{180}{degrees}.
As a consequence, the total mode excitation at the receiver can be written as
\begin{equation}\label{eq:wave-superposition}
\begin{split}
\tilde{B}_{\lambda}(\omega) &= \tilde{B}^+_{\lambda} (\omega ) + {B}^-_{\lambda} (\omega ) e^{i\pi}\\
&= B^+_{\lambda} (\omega ) - B^-_{\lambda} (\omega ).
\end{split}
\end{equation}
Using the definition of $\tilde{B}^\pm_{\lambda} (\omega )$ given by Eq. \eqref{eq:spectral-amplitude-zprop}, we then have
\begin{equation}\label{eq:B-interference}
\tilde{B}_\lambda (\omega) = A^+_\lambda(\omega )e^{ik_\lambda z_r}-A^-_\lambda(\omega ) e^{ik_\lambda (2|z_s - z_t|+z_r)},
\end{equation}
where the expression is being evaluated at the receiver's position, $z_r$, and $z_s$ and $z_t$ are the positions of the reflector and the trap center respectively.
The power spectral density then follows by using Eq. \eqref{eq:power-spectral-density},
\begin{equation}\label{eq:power-short}
\begin{split}
P_\lambda(\omega) = 4 P_{0,\lambda} & \sum _{n=-\infty} ^{\infty} \left\vert a_n \left(\frac{\Omega_0+n\Omega_a}{v_{p.\lambda}}\right)\right\vert^2 \\
&\cos^2\left[(z_t-z_s)\frac{\Omega_0+n\Omega_a}{v_{p.\lambda}}\right]\\
&\left[\delta(\omega - (\Omega_0+n\Omega_a)) + \delta(\omega + \Omega_0+n\Omega_a)\right].
\end{split}
\end{equation}
Here we have assumed that the trap is symmetric, in which case $a_n(-k)$ can be written in terms of $a_n(k)$ as in Eq. \eqref{eq:symmetric-coef-square} (see Appx. \ref{symmetric-trap}).

This power spectrum still has a comb structure, similar to the one in the absence of a reflector at the end of the waveguide.
However, the amplitude of each peak is now modulated with an extra $\cos^2$ factor, which depends on the distance between the reflector and trap center, $z_t-z_s$.
Therefore, while the introduction of a reflector increases the total power collected by the receiver, it also introduces a frequency-dependent amplitude for each peak in the power spectrum.


\section{Trapping Geometries}
\label{sec:trapping-geometries}

In Sec. \ref{sec:radiation-trapped-electron}, we built the foundation for calculating the CRES signal's spectral features.
From the obtained equations, it is clear that it is impossible to extract a simple analytical solution that is valid and usable for every trap configuration.
Therefore, in this section we describe a step-by-step procedure to obtain the spectral properties of a CRES signal.
We will then apply this procedure to two simple and useful trap geometries, enabling us to derive numerical solutions for more complicated geometries following these steps:

\begin{itemize}

    \item An appropriate field approximation $B(z)$ must be found.
    In some cases, where the expression of the exact magnetic field is complex, one can consider using a piecewise approximation of the field.

    \item With the assumed field profile, the electron's equation of axial motion, Eq. \eqref{totenergy}, can be solved.
    Since the effective potential in this equation depends only on the axial position of the electron, we can find a general solution,
\begin{equation}
    t = \int_{z_0(0)}^{z_0(t)} \frac{\mathrm{d}z'}{\sqrt{\frac{2}{m}(K_e - \mu B(z'))}}.
\end{equation}

    \item Once the axial motion of the electron is calculated, the axial frequency follows.
    For the special case of a symmetric trap, we find
\begin{equation}
    \Omega_a ^ {-1} = \frac{2}{\pi} \int_0^{z_{max}} \frac{\mathrm{d}z}{\sqrt{\frac{2}{m}(E_0 - \mu B(z))}}.
\end{equation}

    \item Once the axial position of the electron is found at any given time, the value of magnetic field experienced by the electron at that time, $B(t)$, follows.
    Finally, the cyclotron phase, Eq. \eqref{eq:cyclotron-phase}, is found to be
\begin{equation}
    \Phi_c (t) = \int_0^{t} \frac{eB(t')}{\gamma m_e} \mathrm{d}t'.
\end{equation}

    \item To find the power in each peak, the Fourier coefficients introduced in Eq. \eqref{eq:fourier-series} should be determined by
\begin{equation}
    a_n = \frac{1}{T_a}\int_0^{T_a} e^{i(\Phi_c(t)+k_\lambda z(t))} e^{-i(\Omega_0 + n \Omega_a)t} \mathrm{d}t ,
\end{equation}
in which $\Omega_0$ is the average cyclotron frequency given by
\begin{equation}
    \Omega_0 = \frac{\Phi_c(T_a)}{T_a}.
\end{equation}

    \item The power in each peak of the spectrum can be determined, using Eq. \eqref{eq:sideband-power}, to be
\begin{equation}\label{eq:power-combine}
    P_n = P_{0,\lambda} |a_n|^2.
\end{equation}

    \item Finally, the total power radiated by the electron can be calculated by summing over the power of all peaks.
    This power will define the slopes of tracks in Sec. \ref{sec:spectral-features}.
\end{itemize}

\subsection{Power spectrum in a ``harmonic trap"}

\begin{figure}
\includegraphics[width=0.48\textwidth]{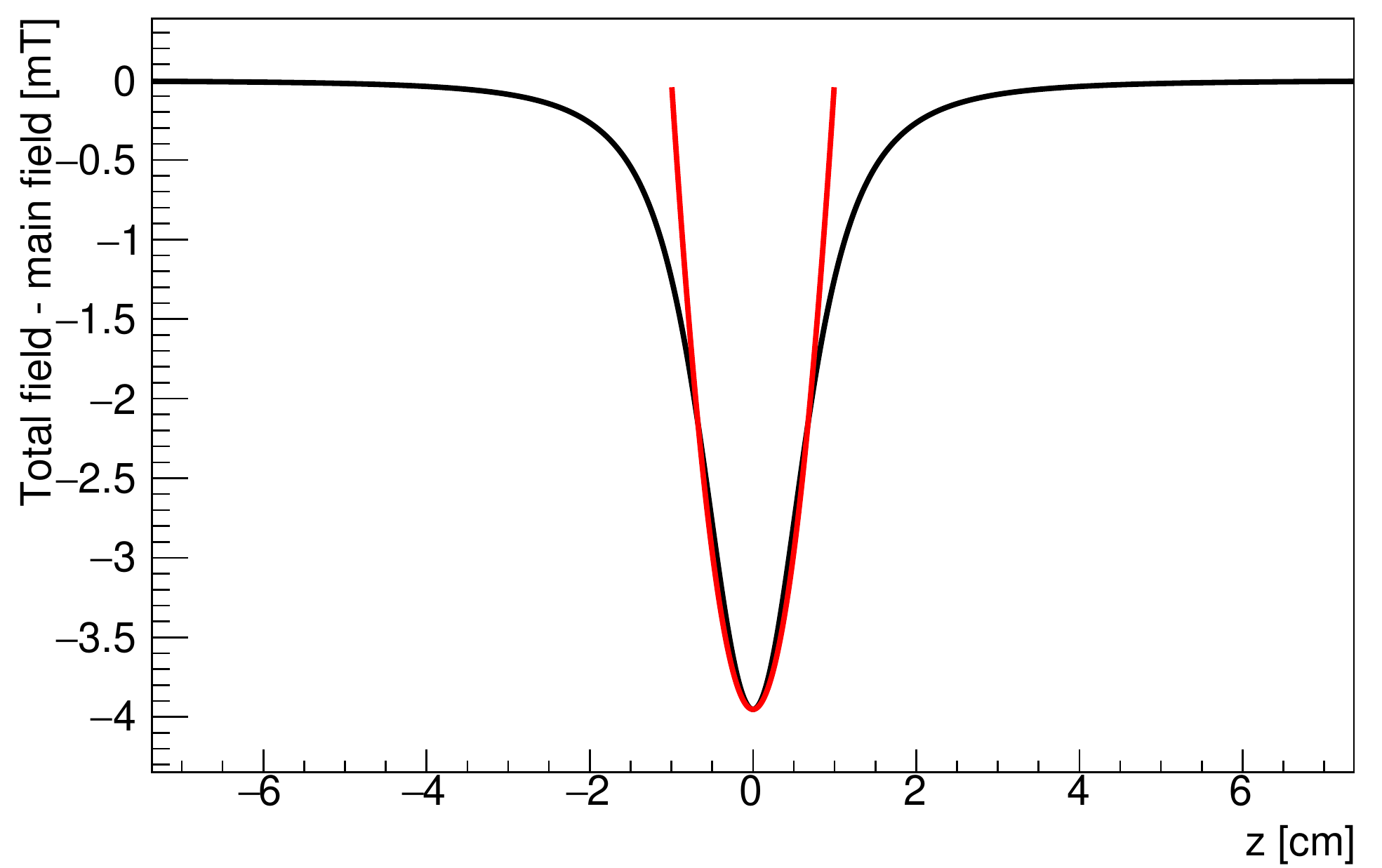}
\caption{\label{fig:harmonic-trap} The on-axis magnetic field profile of a ``harmonic'' trap (black line), generated by a single coil, and the corresponding approximation given by Eq. \eqref{eq:harmonic-approximation} with $L_0 =$ 20~cm (red line).
}
\end{figure}

The simplest magnetic bottle is realized with a single trapping coil producing a field anti-parallel to a background field.
This geometry can be approximated as a purely axial field with parabolic $z$ dependence as represented by Fig. \ref{fig:harmonic-trap}.
It can be described by
\begin{equation}\label{eq:harmonic-approximation}
B_z(z) = B_0 \left(1+\frac{z^2}{L_0^2}\right) ,
\end{equation}
in which $L_0$ is the characteristic length of the trap.
Note that this approximation is accurate for trapped electrons with large pitch angle values that cannot travel to high field regions.

For the harmonic field approximation, electrons undergo simple harmonic motion in the axial direction,
\begin{equation}\label{eq:harmonic-motion}
\begin{split}
&z(t) = z_{\mathrm{max}} \sin(\Omega_a t),
\end{split}
\end{equation}
in which the axial frequency is determined by the axial velocity at the trap minimum,
\begin{equation}\label{eq:axialfreq-harmonictrap}
\Omega_a = \frac{v_0 \sin\theta_{\mathrm{bot}}}{L_0},
\end{equation}
and the maximum displacement for the electron is $z_{\mathrm{max}}~=~L_0 \cot\theta _{\mathrm{bot}}$.

\begin{comment}
\subsubsection{*Cyclotron phase and its expansion in Fourier series}
\end{comment}

The magnetic field seen by the electron as a function of time is
\begin{equation}
B_z(t) = B_0\left(1+\frac{z_{\mathrm{max}}^2}{2L_0^2}-\frac{z_{\mathrm{max}}^2}{2L_0^2}\cos(2\Omega_a t)\right).
\end{equation}
The cyclotron frequency Eq. \eqref{eqn:cyclotron} of a trapped electron,
\begin{equation}
\begin{split}
\Omega_c(t) = & \frac{eB_0}{\gamma m_e} \left(1+\frac{z_{\mathrm{max}}^2}{2L_0^2} - \frac{z_{\mathrm{max}}^2}{2L_0^2}\cos(2\Omega_a t)\right) ,
\end{split}
\end{equation}
follows.
The last term describes the modulation in frequency and the first two terms determine the average cyclotron frequency
\begin{equation}\label{eq5.7}
\Omega_0 = \frac{eB_0}{\gamma m_e} \left(1+\frac{z_{\mathrm{max}}^2}{2L_0^2}\right).
\end{equation}
The cyclotron phase, which can then be found by integrating over the cyclotron frequency, is
\begin{equation}\label{eq5.6}
\Phi_c(t) = \Omega_0 t + q \sin(2\Omega_a t ) ,
\end{equation}
in which the magnitude of the modulation is
\begin{equation}\label{eq5.8}
q = -\frac{eB_0}{\gamma m_e} \frac{z_{\mathrm{max}}^2}{4L_0^2 \Omega_a}.
\end{equation}

\begin{figure}
\includegraphics[width=0.48\textwidth]{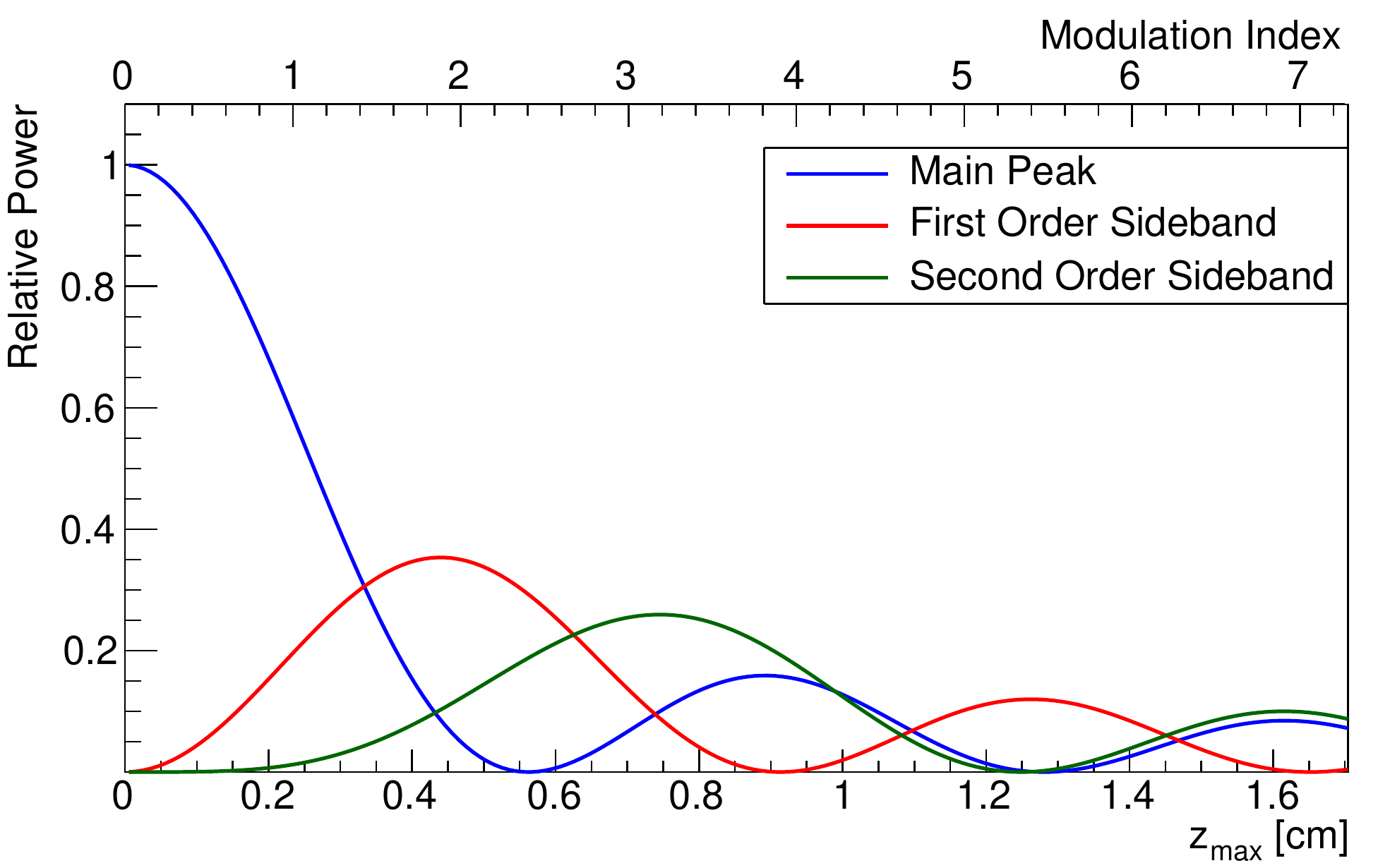}
\caption{\label{fig:sideband_magnitude} Relative magnitudes of the sidebands in a harmonic trap as a function of the maximum axial travel, $z_{\mathrm{max}}$, of a trapped \SI{30}{keV} electron.
Here we consider an ideal harmonic trap as described in Eq. \eqref{eq:harmonic-approximation}, with a background field of 1~T and an $L_0$ of 20~cm.
No reflection effect is taken into account.}
\end{figure}

To find the power spectrum of the electron's radiation, Fourier coefficients in Eq. \eqref{fourier-coefficient} are needed, and can be calculated using the Jacobi-Anger expansion given by
\begin{equation}\label{eq:harmonic-expansion}
\begin{split}
e^{i\Phi_c(t)+ ik_{\lambda}z_0(t)} & = e^{i\left(\Omega_0 t + q \sin(2\Omega_a t)+k_{\lambda} z_{\mathrm{max}} \sin(\Omega_a t)\right)} \\
& = \sum_{m,p=-\infty} ^{\infty} J_m(q)J_p(k_{\lambda}z_{\mathrm{max}}) e^{i\left(\Omega_0+(2m+p)\Omega_a \right)t} ,
\end{split}
\end{equation}
where $J_n$ is the $n^\textrm{th}$ Bessel function of the first kind.
Therefore, the power for each harmonic can be found from Eq.~\eqref{eq:power-combine} by squaring
\begin{equation}
a_n(k_\lambda) = \sum_{m=-\infty} ^{\infty} J_m(q)J_{n-2m}(k_{\lambda}z_{\mathrm{max}})
\end{equation}
and using the appropriate $P_{0,\lambda}$ as found in section \ref{sec:power-waveguide-geometries}.
Let us note that these coefficients, $a_n$, corresponds to the coefficients $\alpha _m$ and $\beta_m$ defined by Eq. \eqref{eq:bathtub-first-expansion} and Eq. \eqref{eq:bathtub-second-expansion}.
This result matches well with our original intuition because the modulation is harmonic, with a modulation index of $q$ for the magnetic field induced modulation, and a modulation index of $k_\lambda z_{\mathrm{max}}$ for the Doppler shift induced modulation.
The relative magnitude of the main peak and sideband powers for typical parameters are shown in Fig.~\ref{fig:sideband_magnitude}.

From Eq. \eqref{eq:trapped-pitch-angles}, a \SI{4}{mT} deep trap in a \SI{1}{T} background magnetic field can trap electrons with pitch angles as small as \SI{86}{degrees}.
In this case, the magnitude of the modulation of the magnetic field experienced by the electron, $q$, will be smaller than 0.6, while the Doppler effect's modulation, $k_\lambda z_{max}$, can be as large as 10.5.
Therefore, $J_m(q)$ can be approximated with $\delta_{m0}$.
In this case, the power spectrum will be simplified to

\begin{equation}\label{eq:simplified-power}
\begin{split}
{\tilde{P}}^\pm_\lambda(\omega) & = P_{0,\lambda} \sum _{n=-\infty} ^{\infty} J_n^2(k_\lambda z_{max})  \\
&\times \left[\delta(\omega - (\Omega_0+n\Omega_a)) + \delta(\omega + \Omega_0+n\Omega_a)\right] .
\end{split}
\end{equation}
This approximation works well for shallow traps in which $\frac{\Delta B}{B} < 0.002$.
\begin{comment}
\subsubsection{*Slope}
\end{comment}

\subsection{Power spectrum in a ``bathtub trap"}

\begin{comment}
\subsubsection{*Field approximation}
\end{comment}

\begin{figure}
\includegraphics[width=0.48\textwidth]{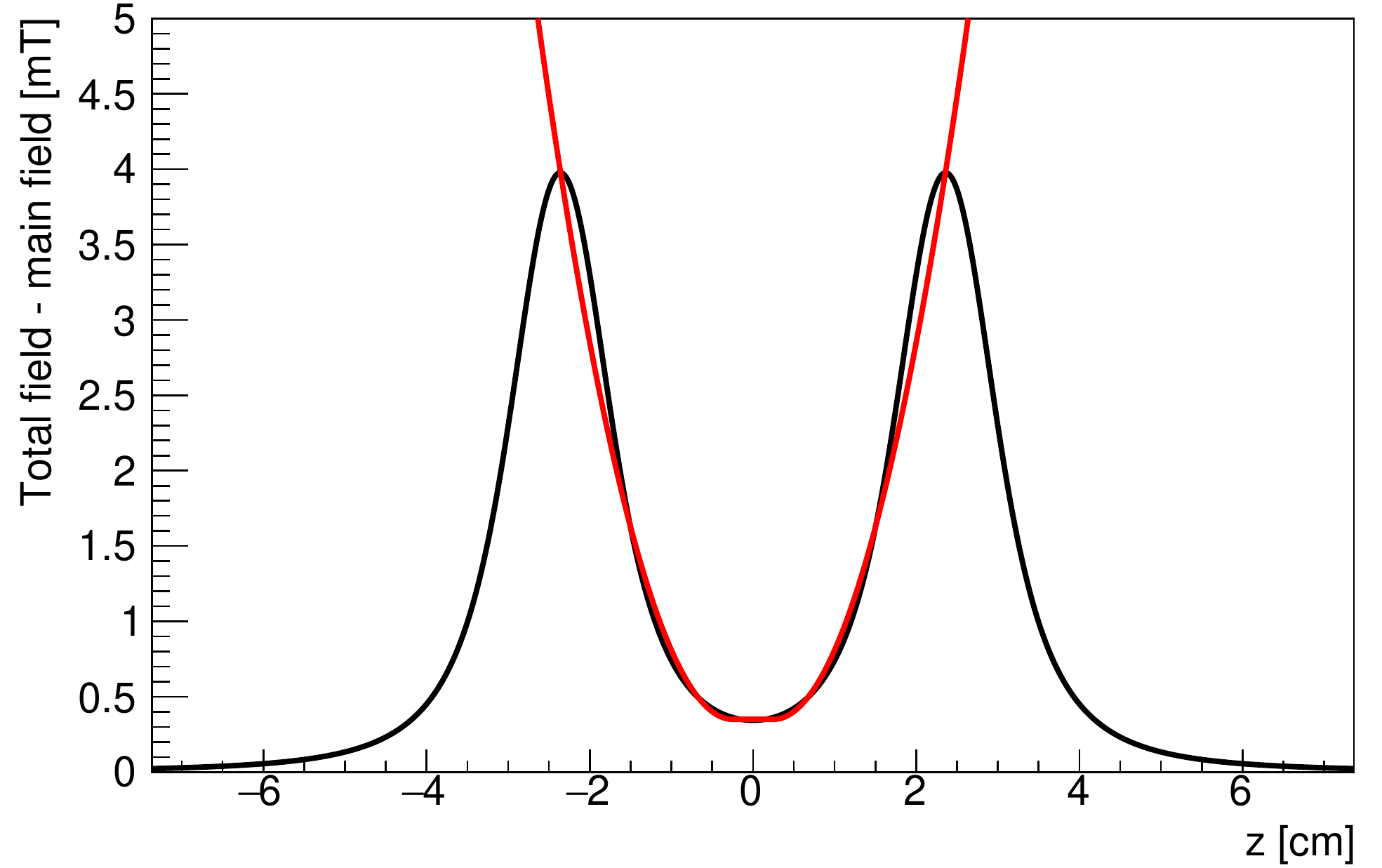}
\caption{\label{fig:bathtub-trap} Magnetic field profile generated by two coils separated by 5~cm, forming a ``bathtub'' shape (black line), and the corresponding approximation given by Eq. \eqref{bath-field} with $L_0$ = \SI{35}{cm} and $L_1$ = \SI{0.5}{cm}(red line).
}
\end{figure}

The harmonic trap described previously has a limited trapping volume.
A ``bathtub trap,'' generated using two coils, includes a wide flat region to extend the trapping volume.
This field geometry is depicted in Fig.~\ref{fig:bathtub-trap}.
In this case, we approximate the field as a region of constant magnetic field between two half parabolas given piecewise by
\begin{equation}\label{bath-field}
B_z(z) = \left\{ \begin{array}{ll}
 B_0\left(1+\frac{\left(z+L_1/2\right)^2}{L_0^2}\right) & z<-\frac{L_1}{2} \\
 B_0 & -\frac{L_1}{2}<z<\frac{L_1}{2} \\
 B_0\left(1+\frac{\left(z-L_1/2\right)^2}{L_0^2}\right) & \frac{L_1}{2}<z
       \end{array} \right. ,
\end{equation}
in which $L_0$ is a measure of field gradient in the curved region and $L_1$ is the width of the flat region.

\begin{comment}
\subsubsection{*Electron motion in the trap}
\end{comment}

For convenience, here we define $t_0$ to be the time when the electron first enters the flat region from the curved region with negative $z$ and we define $t_2$ to be the time one half-period later when it enters the flat region from the opposite direction.
This field configuration results in constant velocity motion when the electron is in the flat region, from $t_0 = 0$ to $t_1 =\frac{L_1}{v_{0}\cos \theta_{\mathrm{bot}}}$, and half harmonic motion at the two ends, for $t_2-t_1 = \frac{\pi}{\omega_a}$, in which the angular frequency $\omega _a$ of the half harmonic motion is defined by $\omega _a = \frac{v_{0}\sin \theta_{\mathrm{bot}}}{L_0}$.
The period of axial motion is then $T=2t_2=\frac{2\pi}{\Omega _a}$, which means that the frequency of the axial motion is
\begin{equation}\label{eq:axialfreq-bathtub}
\Omega_a = \frac{2\pi}{\frac{2L_1}{v_{0}\cos \theta_{\mathrm{bot}}}+\frac{2\pi L_0}{v_{0}\sin \theta_{\mathrm{bot}}}} = \omega _a \left(1+ \frac{L_1}{\pi L_0} \tan \theta_{\mathrm{bot}} \right)^{-1} .
\end{equation}

The equation of axial motion for an electron is thus
\begin{equation} \label{bath-motion}
z(t) = \left\{ \begin{array}{ll}
 v_{z0}t-\frac{L_1}{2} & 0 < t<t_1\\
 z_{\mathrm{max}} \sin [\omega_a (t-t_1)]+\frac{L_1}{2} & 0 < t_1<t<t_2\\
 - v_{z0}(t-t_2)+\frac{L_1}{2} & t_2<t<t_3\equiv t_1+t_2 \\
 - z_{\mathrm{max}} \sin [\omega_a (t-t_3)]-\frac{L_1}{2} & t_3<t<T
       \end{array} \right. ,
\end{equation}
with $z_{\mathrm{max}} = L_0 \cot\theta _{\mathrm{bot}}$ being the maximum displacement for the electron into the harmonic potential.

\begin{comment}
\subsubsection{*Cyclotron phase and its expansion in Fourier series}
\end{comment}

Using this equation and magnetic field configuration from Eq. \eqref{bath-field}, the magnetic field seen by the electron as a function of time is
\begin{equation}
\begin{resize}
B_z(t) = \left\{ \begin{array}{ll}
 B_0 & 0 < t<t_1  \\
 B_0 (1+ \frac{z_{\mathrm{max}}^2}{2L_0^2} - \frac{z_{\mathrm{max}}^2}{2L_0^2} \cos [2\omega_a (t-t_1)]) & t_1<t<t_2 \\
 B_0 & t_2<t<t_3 \\
 B_0 (1+ \frac{z_{\mathrm{max}}^2}{2L_0^2} - \frac{z_{\mathrm{max}}^2}{2L_0^2} \cos [2\omega_a (t-t_3)]) & t_3<t<T
       \end{array} \right. .
\end{resize}
\end{equation}
The cyclotron frequency of the electron is therefore
\begin{equation} \label{bath-freq}
\begin{resize}
\Omega_c(t)=\frac{eB_0}{\gamma m_e} \left\{ \begin{array}{ll}
 1 & 0 < t<t_1  \\
 1+\frac{z_{\mathrm{max}}^2}{2L_0^2}-\frac{z_{\mathrm{max}}^2}{2L_0^2} \cos [2\omega_a (t-t_1)] & t_1<t<t_2 \\
 1 & t_2<t<t_3 \\
 1+\frac{z_{\mathrm{max}}^2}{2L_0^2}- \frac{z_{\mathrm{max}}^2}{2L_0^2} \cos [2\omega_a (t-t_3)] & t_3<t<T
  \end{array} \right. .
\end{resize}
\end{equation}\\
and the average frequency of cyclotron radiation is
\begin{equation}\label{bathtub-start-frequency}
\Omega_0 = \frac{eB_0}{\gamma m_e} \left(1+\frac{z_{\mathrm{max}}^2}{2L_0^2} \left(1+ \frac{L_1}{\pi L_0} \tan \theta_{\mathrm{bot}} \right)^{-1}\right) .
\end{equation}

\begin{comment}
\subsubsection{*Power spectrum}
\end{comment}

The detailed calculation of the coefficients $a_n$, which are used to calculate the power, can be found in Appx. \ref{sec:bathtub-calculation}.

\begin{comment}
\subsubsection{*Slope}
\end{comment}


\section{Spectral Features in Cyclotron Radiation Emission Spectroscopy}
\label{sec:spectral-features}

\begin{comment}
\subsection{*Introduction}
\end{comment}

In this section we identify the features required for reconstructing the kinematics of an electron in a CRES experiment, based on the relationships in previous sections. We also develop a common terminology for these features.

\begin{comment}
\subsection{*Tracks}
\end{comment}

\begin{figure}
\includegraphics[width=0.48\textwidth]{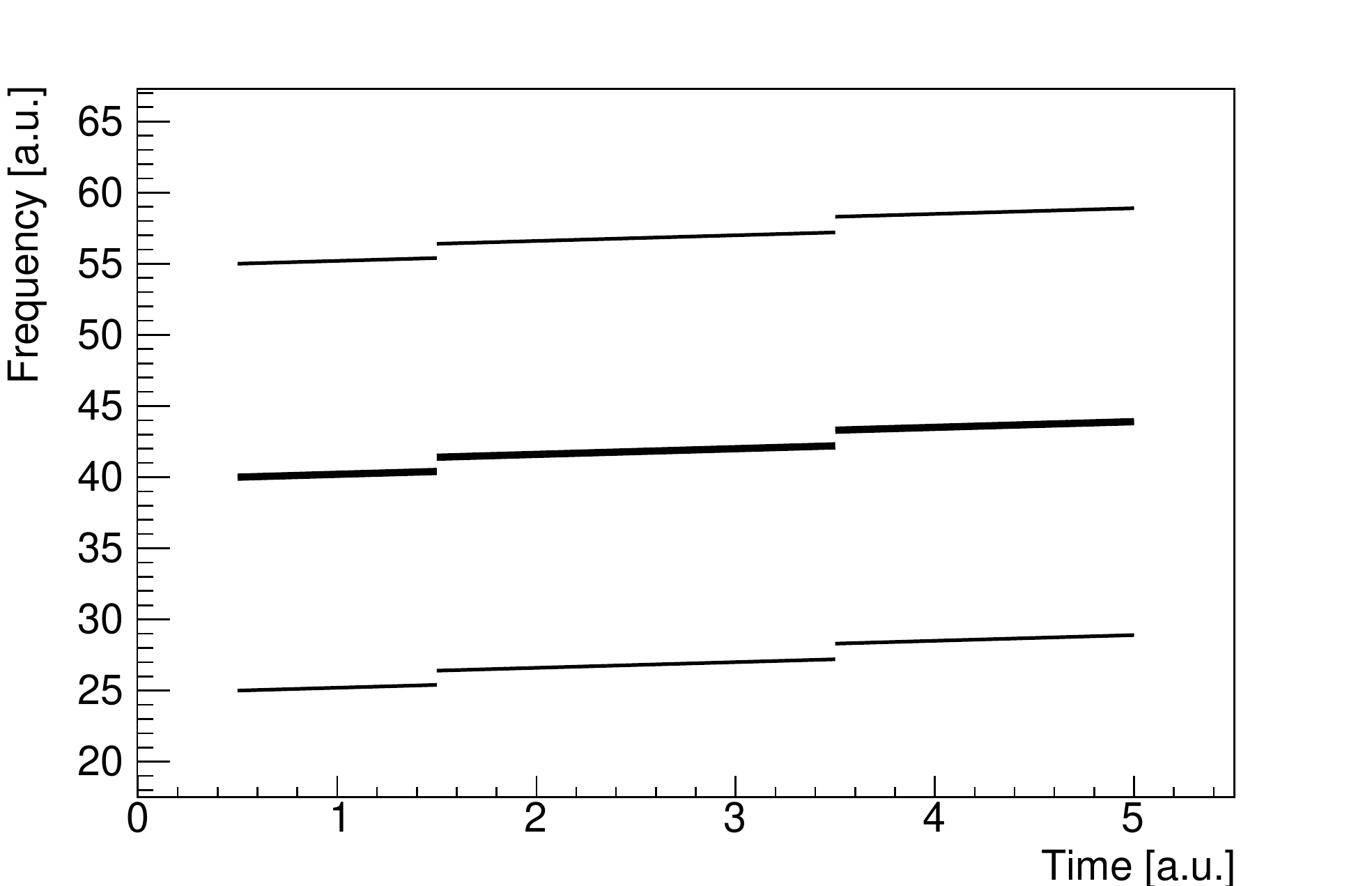}
\caption{\label{fig:waterfall} Schematic of the power (represented by line width), as a function of time and frequency in the absence of a waveguide reflector.
The main track and first order sidebands are shown.
Sudden losses of energy (and thus increases of frequency), induced by collisions with background gas particles, happen at $1.5~\mathrm{ms}$ and $3.5~\mathrm{ms}$.}
\end{figure}

The power spectrum of the signal generated by an electron possesses a comb structure given by Eq. \eqref{eq:power_comb_structure}.
If we represent the power spectrum as a function of time in a spectrogram, the excess of power forms connected structures that we call \textit{tracks}.
Fig. \ref{fig:waterfall} represents the tracks coming from the comb structure of the spectrum.
The track at the average cyclotron frequency, given by Eq. \eqref{eqn:approx_instantaneous_cyclotron}, is called the \textit{main track}.
As the electron radiates energy, the cyclotron frequency increases, causing the tracks to have a positive slope.
For any given trap configuration, the track's slope, S, is proportional to the power radiated into both propagating and non-propagating modes.
According to Eq. \eqref{eqn:approx_instantaneous_cyclotron} this relation can be written as
\begin{equation}
S = \frac{\Omega_c}{m_e c^2 + K_0} P.
\end{equation}
Electrons can scatter off a molecule of the residual gas in the waveguide, causing abrupt energy losses, changes of pitch angle, and breaks in the observed tracks.

\begin{comment}
\subsection{*Sidebands}
\end{comment}

The tracks parallel to the main tracks we call {\it sidebands}.
These tracks are located at multiples of the axial frequency, $f_a = \frac{\Omega _a}{2\pi}$, away from the main track; the {\it order} of a sideband corresponds to this multiplicity.
As long as we only consider time intervals short enough that the power radiated does not significantly change the axial frequency, sidebands will appear parallel to the main track.
Eq. \eqref{eq:axialfreq-harmonictrap} and Eq. \eqref{eq:axialfreq-bathtub} show how the axial frequency, measured from the frequency-distance to sidebands, can be used to relate the pitch angle and kinetic energy of an electron in a harmonic or bathtub traps.

\begin{comment}
\subsection{*Track Power and Slope}
\end{comment}

\begin{figure}
\includegraphics[width=0.48\textwidth]{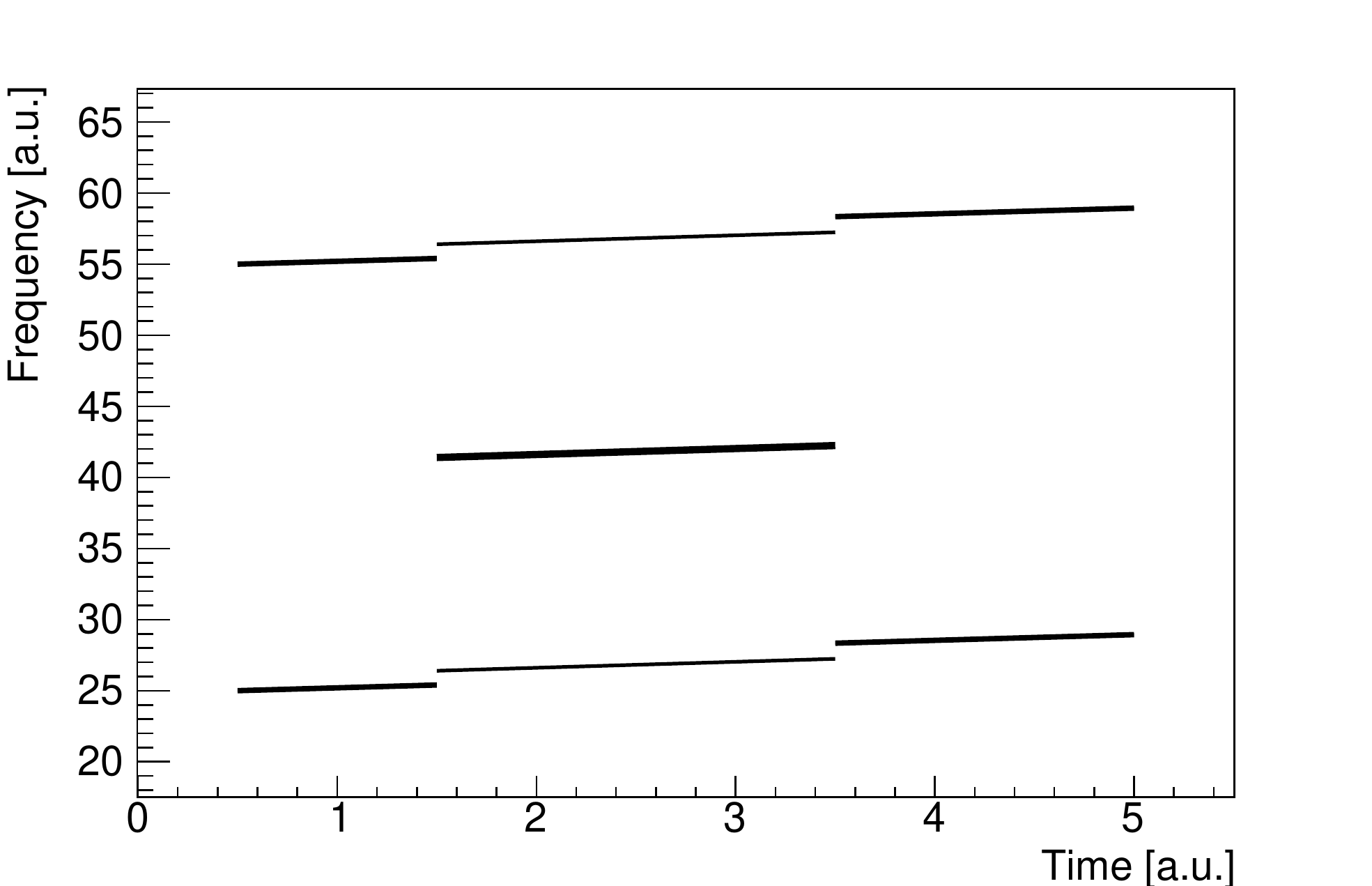}
\caption{\label{fig:app-disapp-sidebands} Schematic of the power (represented by line width) as a function of time and frequency in the presence of a waveguide reflector.
The main track appears and disappears as the kinematic parameters (such as the pitch angle) change as a result of collisions with background gas particles. }
\end{figure}

The distribution of power between a main track and its sidebands depends on the electron's energy and pitch angle.
In the presence of a reflector on one end of the waveguide, as described in Section \ref{sec:effect-experimental-config}, the distance between the trap and the reflector will also impact the power distribution as shown in Fig.~\ref{fig:app-disapp-sidebands}.
In realistic experiments, this is further complicated as tracks with suppressed power will be indistinguishable from noise.


\section{Extraction of Kinematic Parameters from a Measured Spectrum}
\label{sec:extraction-of-kinematic-parameters}

The previous sections show that the primary observable parameters of a CRES signal are the frequency of the main track, the frequency separating the sidebands, the power in both the main track and sidebands, and the slope of the main track.

For a given configuration of trapping field and waveguide, these parameters are completely determined by the electron's kinetic energy and pitch angle.
However, the converse is not in general true. The axial frequency in a real magnetic trap is double valued with respect to the pitch angle whenever the floor of the trap is flatter than harmonic.
This is because the axial frequency is relatively low both for small amplitudes and for amplitudes that almost eject the electron over the trap-field maxima, and it reaches a broad maximum for intermediate amplitudes.
Other ambiguities arise when resonant structures such as those described in Sec. \ref{sec:effect-experimental-config} cause the slope to have multiple values.
These ambiguities can be mitigated at the design stage and by making use of all the available information in the signal.
We will now give a concrete example of predicting the observable parameters from a particular trapping field, and then speculate on the observations needed to reconstruct the electron's initial kinetic energy.

\begin{figure}[htb!]
\includegraphics[width=0.48\textwidth]{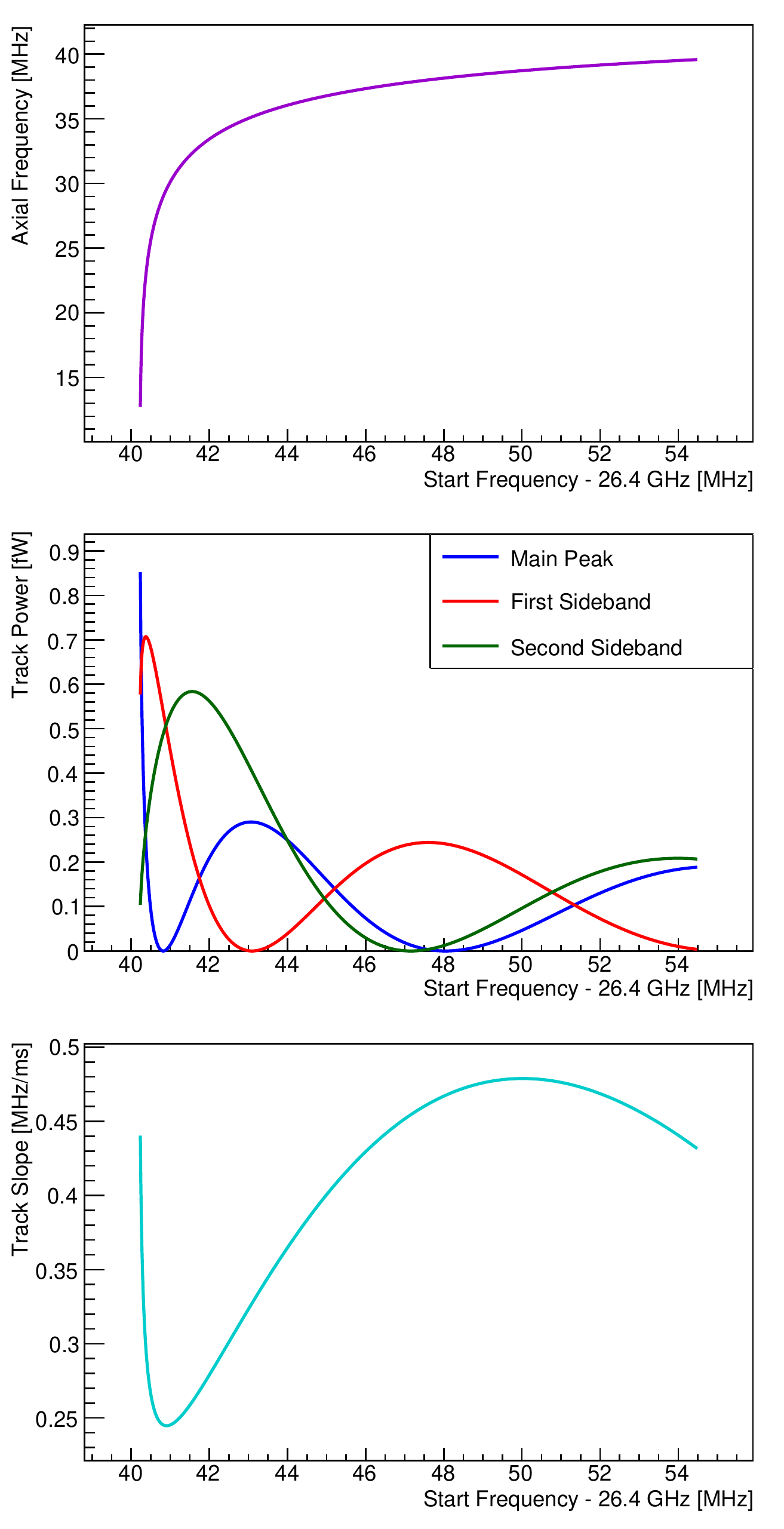}
\begin{minipage}[b]{0.48\textwidth}
\caption{\label{fig:power_Slope} Spectral features of the CRES signal for \SI{30}{keV} electrons with different values of pitch angle and single radial position at $\rho =$ 0~cm, trapped in an ideal baththub trap described by Eq.\eqref{bath-field} with $L_0 =$ \SI{35}{cm} and $L_1 =$ \SI{0.5}{cm}, in a \SI{1}{T} background field including the effect of a short.
An electron with a pitch angle of 90 degrees has a start frequency of \SI{26.44}{GHz} while electrons with lower pitch angles are subject to pitch angle effects which increase their start frequencies.
This shift can systematically affect energy measurements in CRES experiments.
The above plots illustrate how different measurable quantities in a CRES experiment can be used to correct for this frequency shift.}
\end{minipage}
\end{figure}

For our example in Fig. \ref{fig:power_Slope}, we will use a bathtub trap with an $L_0$ of \SI{35}{cm} and an $L_1$ of \SI{0.5}{cm} in a \SI{1.07}{cm} wide rectangular waveguide.
We consider a short on one end of the waveguide, a distance \SI{0.6}{cm} away from the trap center, and a \SI{1}{T} background magnetic field.
We will examine predicted signals from electrons with \SI{30}{keV} of kinetic energy and with different pitch angles.
We find the power in the $n^\mathrm{th}$ harmonic for this situation using Eq.~\eqref{eq:power-short}, which includes the short, using the power from Eq.~\eqref{eq3.14}, which is for the rectangular waveguide.
The Fourier coefficients for the bathtub trap are found in Appx. \ref{sec:bathtub-calculation}.
Therefore we have
\begin{equation}\label{eq:power-short-final}
\begin{resize}
\begin{split}
P_\lambda(\omega) & = \frac{Z_{{10}}e^2v_0^2}{\pi wh} \cos ^2 \left(\frac{\pi x_c}{w}\right) \\
& \sum _{n=-\infty} ^{\infty} \left\vert a_n \left(k_\lambda\right)\right\vert^2 \cos^2\left[(z_t-z_s)k_\lambda \right] \\
&\left[\delta(\omega - (\Omega_0+n\Omega_a)) + \delta(\omega + \Omega_0+n\Omega_a)\right].
\end{split}
\end{resize}
\end{equation}

\begin{figure}[tb]
\includegraphics[width=0.48\textwidth]{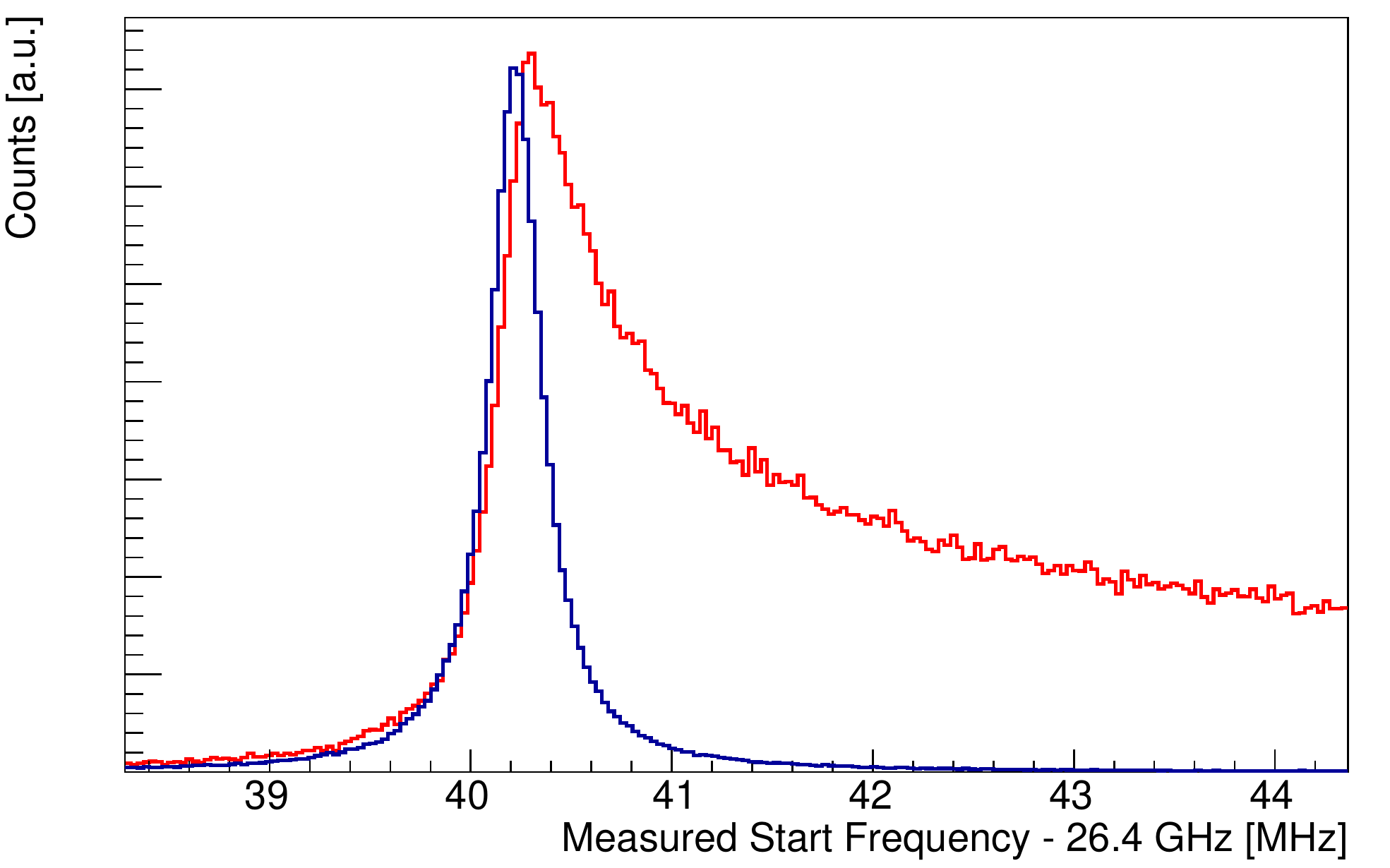}
\caption{\label{fig:lineshape-distortion} Simulation of the energy spectrum of  electrons sampled from a $60~\mathrm{eV}$ wide Lorentzian, centered at $30~\mathrm{keV}$, in a $1~\mathrm{T}$ background magnetic field.
In blue, the spectrum of the extracted start frequencies for an idealized case where the magnetic field is flat and all electrons have a \SI{90}{degrees} pitch angle.
In red, the actual lineshape when the electrons have an isotropic momentum distribution and are confined in a \SI{4}{mT} deep ideal harmonic trap as given in Eq. \eqref{eq:harmonic-approximation}.
The blue histogram is scaled down, so it can be compared with the red one.}
\end{figure}

Of the observables, the \textit{start frequency} of the main track is the most strongly related to the electron's kinetic energy.
The cyclotron frequency of a \SI{30}{keV} electron, with a pitch angle of \SI{90}{degrees} at the center of the trap, is \SI{26.44}{GHz}.
However, this frequency is increased by the electron's pitch angle as described in Sec. \ref{sec:axial-motion}.
The distortion of the distribution of main track frequencies by pitch angle is shown in Fig. \ref{fig:lineshape-distortion}.

The other signal parameters can be used to correct for the pitch angle effect and recover the true kinetic energy of the electron.
Decreasing the pitch angle will simultaneously increase the start frequency and effects the other parameters discussed above.
In principle, only a subset of the parameters are needed to find the pitch angle and recover the correct energy.

A sufficiently precise measurement of the axial frequency alone can be used to correct the main track frequency, yielding the cyclotron frequency at the center of the trap.
However, extraction of the axial frequency is possible only for pitch angles for which there are at least two visible tracks above the noise level.

In other cases, other parameters must be used.
Track power carries valuable information, though typically power measurements in CRES experiments are less precise than frequency measurements, and may not be possible if the noise level is high.
Furthermore, the power is double-valued for non-shallow trap geometries that can trap electrons with smaller pitch angles.
Determining a track's slope is a frequency measurement, measurable even if the main track or sidebands are absent, and therefore is a most reliable parameter for correction.
However, the slope is double-valued for this example.
The precise algorithm for combining the parameters to achieve a high resolution energy measurement will, therefore, depend on the particular geometry and signal-to-noise of the CRES experiment.

\section{Conclusion}

We have found that electrons in a CRES experiment undergo nontrivial but predictable motion within a magnetic bottle, and this motion affects the detected cyclotron signal.
We identified the carrier and sideband structure of the signal, and have shown that these features encode the entirety of the electron's kinematic parameters.
Following the results derived here, a sufficiently precise measurement of these features should allow complete reconstruction of the electron's kinetic energy, which is necessary for proposed CRES experiments to achieve their desired sensitivity.
In fact, the measurable features over-constrain the kinematic parameters and may be able to calibrate some of the detector configuration as well.
Notably, we point out that for configurations where the electron undergoes axial motion larger than a half wavelength of cyclotron radiation, the modulation is such that detection and interpretation of sidebands is necessary to detect all trapped electrons.

The practicalities of signal detection and reconstruction will depend on the particular apparatus design and detection scheme, in particular the signal-to-noise ratio of the sidebands, and we leave a discussion of the precise reconstruction algorithm and ultimate resolution to future work.

\section{Acknowledgments}

This material is based upon work supported by the following sources: the U.S. Department of Energy Office of Science, Office of Nuclear Physics, under Award No.~DE~SC0014130 to UCSB, under Award No.~DE-SC0011091 to MIT, under the Early Career Research Program to Pacific Northwest National Laboratory (PNNL), a multiprogram national laboratory operated by Battelle for the U.S. Department of Energy under Contract No.~DE-AC05-76RL01830, under Award No.~DE-FG02-97ER41020 to the University of Washington, and under Award No.~DE-SC0012654 to Yale University; the National Science Foundation under Award Nos.~1205100 and 1505678 to MIT; Lab-Directed Research and Development at LLNL (18-ERD-028), Prepared by LLNL under Contract DE-AC52-07NA27344; the Massachusetts Institute of Technology (MIT) Wade Fellowship; the Laboratory Directed Research and Development Program at PNNL; the University of Washington Royalty Research Foundation. A portion of the research was performed using Research Computing at Pacific Northwest National Laboratory. We further acknowledge support from Yale University, the PRISMA Cluster of Excellence at the University of Mainz, and the KIT Center Elementary Particle and Astroparticle Physics (KCETA) at the Karlsruhe Institute of Technology.

\appendix
\section{Property of the Fourier coefficients in a symmetric trap}
\label{symmetric-trap}

In a trap where the magnetic field distortion is symmetric with respect to the center of the trap,  we expect the same amplitude of radiation to propagate in both directions in the waveguide.
This means that we need to show that
\begin{equation}\label{eq:symmetric-coef-square-initial}
\left|a_n(-k_\lambda)\right|^2 = \left|a_n(k_\lambda)\right|^2.
\end{equation}

Two useful expressions in symmetric traps will assist us in deriving this relation. The first relates to the periodicity of the electron's position, $z_0$, in a symmetric trap, given by
\begin{equation}\label{eq:symmetric-position}
-(z_0(t) - z_t) = z_0\left(t+\frac{T_a}{2}\right)- z_t ,
\end{equation}
in which $z_t$ is the axial position of the center of the trap.
Furthermore, a symmetric trap forces the cyclotron frequency to be periodic, with period equal to half of the axial motion's period.
Therefore the cyclotron phase satisfies
\begin{equation} \label{eq:symmetric-phase}
\Phi_c\left(t+\frac{T_a}{2}\right) = \Omega_0 \frac{T_a}{2} + \Phi_c(t).
\end{equation}

Utilizing Eq. \eqref{eq:symmetric-position} for $\Phi_c(t)$ and Eq. \eqref{eq:symmetric-phase} to rewrite $k_\lambda z_0(t)$, we can write
\begin{equation}
\begin{split}
\Phi_c(t)- k_{\lambda}z_0(t) & = \Phi_c\left(t+\frac{T_a}{2}\right)-\Omega_0 \frac{T_a}{2}\\
&+ k_{\lambda}z_0\left(t+\frac{T_a}{2}\right)-2k_\lambda z_t .
\end{split}
\end{equation}
Therefore we have
\begin{equation}
\begin{split}
e^{i\Phi_c(t)- ik_{\lambda}z_0(t)} & = e^{-i\Omega_0 \frac{T_a}{2}-2ik_\lambda z_t}e^{i\Phi_c(t+\frac{T_a}{2})+ ik_{\lambda}z_0(t+\frac{T_a}{2})} .
\end{split}
\end{equation}
Using Eq. \eqref{eq:fourier-series}, we expand the second exponent to get
\begin{equation}
\begin{split}
e^{i\Phi_c(t)- ik_{\lambda}z_0(t)} & = e^{-2ik_\lambda z_t-i\Omega_0 \frac{T_a}{2}}\sum _{n=-\infty} ^{\infty} a_n(k_{\lambda}) e^{i(\Omega_0+n\Omega_a)(t+\frac{T_a}{2})} \\
& = e^{-2ik_\lambda z_t}\sum _{n=-\infty} ^{\infty} (-1)^n a_n(k_{\lambda}) e^{i(\Omega_0+n\Omega_a)t} .
\end{split}
\end{equation}

By equating the coefficients with those of the first expression in Eq. \eqref{eq:other-fourier-series}, we arrive at the form,

\begin{equation}\label{eq:symmetric-trap-coef}
a_n(-k_\lambda) = (-1)^n  e^{-2ik_\lambda z_t} a_n(k_{\lambda}),
\end{equation}
which is consistent with our expectation of equal power propagating in both directions, since

\begin{equation}\label{eq:symmetric-coef-square}
\left|a_n(-k_\lambda)\right|^2 = \left|a_n(k_\lambda)\right|^2.
\end{equation}

\section{$P_{0,\lambda}$ calculation for two specific waveguide geometries}
\label{sec:power-term}
The power amplitude, $P_{0,\lambda}$, was introduced in Eq. \eqref{p0-def} as a measurement of an electron's coupling to a waveguide mode.
The calculation details for two particularly relevant cases are shown here.

\subsection{Rectangular waveguide $TE_{10}$ mode}
The first example is the fundamental mode of a rectangular waveguide.
For such a waveguide, with $w$ being its longer dimension (defined to be along the $x$ axis) and $h$ the smaller one (along the $y$ axis), the electric field has the form
\begin{equation}\label{eqa.2}
E_y(x) = K \cos\left(\frac{\pi x}{w}\right) \hat{y}.
\end{equation}
Eq. \eqref{eq3.2} can now be used to find the normalization factor, giving
\begin{equation}\label{eqa.3}
\int _{\mathcal{A}} K^2 \cos^2 \left(\frac{\pi x}{w}\right) \mathrm{d}x\mathrm{d}y = 1 \Rightarrow K = \sqrt{\frac{2}{wh}} .
\end{equation}

With the normalized field, the expression for $P_{0,TE10}$ follows from the definition in Eq. \eqref{p0-def} and is found to be
\begin{equation}
\begin{split}
P_{0,TE_{10}} & = \frac{Z_{{10}}e^2v_0^2}{8} \left( \sqrt{\frac{2}{wh}} \cos \left(\frac{\pi x_c}{w}\right)\right)^2 \\
& = \frac{Z_{{10}}e^2v_0^2}{4 wh}\cos^2\left(\frac{\pi x_c}{w}\right)  .
\end{split}
\end{equation}

\subsection{Circular waveguide $TE_{11}$ mode}
The second example to consider is that of a circular waveguide with radius $R$.
The $TE_{11}$ mode has the lowest cutoff frequency in a circular waveguide and the associated wavenumber is $k_c = \frac{1.841}{R}$.
This mode consists of two degenerate modes for which the electric field can be found in \cite{Pozar2004},
\begin{equation} \label{eqa.5}
\begin{split}
& E_{1\rho} (\rho,\phi) = K \frac{-i\omega\mu}{k_c^2 \rho } \cos(\phi)J_1(k_c\rho),\\
& E_{1\phi} (\rho,\phi) = K \frac{i\omega\mu}{k_c} \sin(\phi)J_1'(k_c\rho),\\
& E_{1z} (\rho,\phi) = 0
\end{split}
\end{equation}
and
\begin{equation} \label{eqa.5.1}
\begin{split}
& E_{2\rho} (\rho,\phi) = K' \frac{-i\omega\mu}{k_c^2 \rho } \sin(\phi)J_1(k_c\rho),\\
& E_{2\phi} (\rho,\phi) = K' \frac{i\omega\mu}{k_c} \cos(\phi)J_1'(k_c\rho),\\
& E_{2z} (\rho,\phi) = 0 .
\end{split}
\end{equation}
The same technique is used to find the normalized fields,
\begin{equation}\label{eqa.6}
\begin{split}
1 & = \int _{\mathcal{A}} \left[E_{1\rho}^2(\rho,\phi) + E_{1\phi}^2(\rho,\phi)\right] \rho \mathrm{d}\rho \mathrm{d}\phi  \\
& = - K^2 \pi \frac{\omega^2\mu^2}{2k_c^2}\int_{0}^{R} \left[ \frac{J_1^2(k_c\rho)}{k_c^2 \rho^2}+J_1'^2(k_c\rho) \right] \rho \mathrm{d}\rho .
\end{split}
\end{equation}
Hence the normalization factor can be found to be
\begin{equation}\label{eqa.7}
K = K' = \frac{i k_c}{\omega \mu \sqrt{\pi\alpha}}
\end{equation}
in which
\begin{equation}\label{eqa.8}
\alpha = \int_{0}^{R} \left[\frac{J_1^2(k_c\rho)}{k_c^2 \rho^2}+J_1'^2(k_c\rho)\right] \rho \mathrm{d}\rho .
\end{equation}

The calculation of the coefficients $P_{0,TE_{11}}$  follows the rectangular waveguide calculation with one difference.
That is, to find the power in the waveguide, the two degenerate modes' powers should be added together.
This gives us
\begin{equation}\label{eqa.9}
\begin{split}
P_{0,TE_{11}} & = \frac{Z_{{11}}e^2v_0^2}{8} \left[ E_{1\phi}^2 + E_{1\rho}^2 + E_{2\phi}^2 + E_{2\rho}^2\right] \\
& = \frac{Z_{{11}}e^2 v_0^2}{8\pi\alpha} \left( J_1'^2(k_c \rho_c) + \frac{1}{k_c^2 \rho_c^2} J_1^2(k_c \rho_c)\right) .
\end{split}
\end{equation}

\section{Bathtub Trap Calculation}
\label{sec:bathtub-calculation}

The ``bathtub'' trapping geometry was introduced in Sec. \ref{sec:trapping-geometries}.
Here we present detailed calculations of both the phase and the axial motion Fourier expansion coefficients, defined by Eq. \eqref{eq:bathtub-first-expansion} and Eq. \eqref{eq:bathtub-second-expansion}, respectively.

First, we define the frequency difference between the average cyclotron and the frequency at the bottom of the trap using Eq. \eqref{eq:axialfreq-bathtub},
\begin{equation}\label{eq:bathtub-freq-change}
\begin{split}
\Delta \Omega &\equiv \Omega_0- \frac{eB_0}{\gamma m_e} = \frac{\Omega_c}{2 }\frac{z_{\mathrm{max}}^2}{L_0^2}  \left(1+\frac{L_1}{ \pi L_0}\tan \theta \right)^{-1} \\
 &= \frac{\Omega_c}{2 }\frac{z_{\mathrm{max}}^2}{L_0^2} \frac{\Omega _a}{\omega _a}.
\end{split}
\end{equation}
The perturbation to the average cyclotron phase can be written as
\begin{equation} \label{eqc.6}
\begin{resize}
\begin{split}
&\Phi_c(t)-\Omega_0 t =\\
&\left\{ \begin{array}{ll}
 -\Delta \Omega t & 0 < t <t_1  \\
\Delta \Omega \frac{\omega_a}{\Omega _a} (t-t_1)- \frac{\Delta \Omega }{2\Omega _a}\sin [2 \omega_a (t-t_1)]-\Delta \Omega  t & t_1<t<t_2 \\
 -\Delta \Omega (t-t_2) & t_2<t<t_3 \\
\Delta \Omega \frac{\omega_a}{\Omega _a} (t-t_3)-\frac{\Delta \Omega }{2\Omega _a} \sin [2 \omega_a (t-t_3)]-\Delta \Omega  (t-t_2) & t_3<t<T
       \end{array} \right. .
\end{split}
\end{resize}
\end{equation}

The coefficients, $\alpha_n$, can then be found to be
\begin{equation}
\alpha_n = \frac{1}{T} \int_0^T e^{i\Phi_c(t)-i\Omega_0 t} e^{-in\Omega_a t} \mathrm{d}t .
\end{equation}
The integral can be computed by splitting it into four pieces as
\begin{equation}\label{eq:alpha-parts}
\alpha_n = \frac{1}{T} \left( A_n + B_n + C_n + D_n\right),
\end{equation}
in which
\begin{align}
A_n & = \int_0 ^ {t_1} e^{i\Phi_c(t)-i\Omega_0  t} e^{-in\Omega_a t} \mathrm{d}t \nonumber\\
& = t_1 e^{-i \left(\Delta \Omega + n \Omega_a \right) \frac{t_1}{2}} \sinc\left[\left(\Delta \Omega + n \Omega_a \right) \frac{t_1}{2}\right],\\
B_n & = \int_{t_1} ^ {t_2} e^{i\Phi_c(t)-i\Omega_0  t} e^{-in\Omega_a t} \mathrm{d}t \nonumber\\
& = \frac{\pi}{\omega _a}e^{-i\left(\Delta\Omega + n\Omega_a\right) t_1/2}\sum _{m=-\infty}^{\infty} J_m\left(\frac{\Delta\Omega}{2\Omega_a}\right)e^{-in\frac{\pi}{2}}\nonumber\\
& \sinc \left(\Delta\Omega \frac{t_1}{2} -\frac{n\pi}{2}\frac{\Omega _a}{\omega_a}+m\pi\right),\\
C_n & = \int_{t_2} ^ {t_3} e^{i\Phi_c(t)-i\Omega_0  t} e^{-in\Omega_a t} \mathrm{d}t \nonumber\\
& = (-1)^n A_n,\\
D_n & = \int_{t_3} ^ {T} e^{i\Phi_c(t)-i\Omega_0  t} e^{-in\Omega_a t} \mathrm{d}t \nonumber\\
& = (-1)^n B_n.
\end{align}
Note that for odd values of $n$ the coefficient $\alpha_n$ is zero.

The determination of $\beta _n$ follows in a similar manner.
The electron's equation of motion (Eq. \eqref{bath-motion}) gives
\begin{equation}
\begin{split}
\beta_n &= \frac{1}{T} \int_0^T e^{ik_\lambda z(t)} e^{-in\Omega_a t} \mathrm{d}t\\
&= \frac{1}{T} \left( E_n + F_n + G_n + H_n\right),
\end{split}
\end{equation}
where
\begin{align}
E_n & = \int_{0} ^ {t_1} e^{ik_\lambda z(t)} e^{-in\Omega_a t} \mathrm{d}t \nonumber\\
& = t_1 e^{- i n \Omega_a \frac{t_1}{2}} \sinc\left[\left(k_\lambda v_{z0} - n \Omega_a \right)\frac{t_1}{2}\right],\\
F_n & = \int_{t_1} ^ {t_2} e^{ik_\lambda z(t)} e^{-in\Omega_a t} \mathrm{d}t \nonumber\\
& = e^{ik_\lambda L_1/2}\frac{\pi}{\omega_a}e^{-in\Omega_a t1/2}\nonumber\\
&\sum_{m=-\infty}^{\infty} J_m(k_{\lambda}z_{\mathrm{max}})i^{m-n}\sinc \left( \frac{m\pi}{2}-\frac{n\pi}{2}\frac{\Omega_a}{\omega_a}\right),
\end{align}

\vfill\eject

\begin{align}
G_n & = \int_{t_2} ^ {t_3} e^{ik_\lambda z(t)} e^{-in\Omega_a t} \mathrm{d}t \nonumber\\
& = (-1)^n \ t_1 e^{-i n \Omega_a \frac{t_1}{2}} \sinc\left[\left(k_\lambda v_{z0} + n \Omega_a \right)\frac{t_1}{2}\right],\\
H_n & = \int_{t_3} ^ {T} e^{ik_\lambda z(t)} e^{-in\Omega_a t} \mathrm{d}t \nonumber\\
& =(-1)^n e^{-ik_\lambda L_1/2}\frac{\pi}{\omega_a}e^{-in\Omega_a t1/2} \nonumber\\
&\sum_{m=-\infty}^{\infty} J_m(k_{\lambda}z_{\mathrm{max}}) i^{-m-n}\sinc \left( \frac{m\pi}{2}-\frac{n\pi}{2}\frac{\Omega_a}{\omega_a}\right).
\end{align}

The coefficients $\alpha_n$ and $\beta_n$ can be used to find $a_n$ as defined in Eq. \eqref{fourier-coefficient}. These $a_n$ coefficients are a measure of the relative power in the $n^\mathrm{th}$ peak of the power spectrum, according to Eq. \eqref{eq:power_comb_structure}.

\bibliographystyle{apsrev4-1}
\bibliography{2017_esfahani_rybka.bib}

\end{document}